# Universal gravitation as a complex-dynamical process, renormalised Planckian units, and the spectrum of elementary particles

A.P. KIRILYUK*

Institute of Metal Physics, Kiev, Ukraine 252142

ABSTRACT. The new, complex-dynamical mechanism of the universal gravitation naturally incorporating dynamical quantization, wave-particle duality, and relativity of physically emerging space and time (quant-ph/9902015,16) provides the realistic meaning and fundamentally substantiated modification of the Planckian units of mass, length, and time approaching them closely to the extreme characteristic values observed for already discovered elementary particles. This result suggests the important change of the whole strategy of practical research in high-energy/particle physics, decisively displacing it towards the already attained energy scales and permitting one to exclude the existence of elementary objects in the inexplicably large interval separating the known, practically quite sufficient set of elementary species and the conventional, mechanistically exaggerated values of the Planckian units. This conclusion is supported by the causally complete (physically realistic and mathematically consistent) picture of the fundamental levels of reality derived, without artificial introduction of any structure or 'principle', from the unreduced analysis of the (generic) interaction process between two primal, physically real, but a priori structureless entities, the electromagnetic and gravitational protofields. The naturally emerging phenomenon of universal dynamic redundance (multivaluedness) of any interaction process gives rise to the intrinsically unified hierarchy of unreduced dynamic complexity of the world providing the causally complete, physically transparent origin of elementary particles/fields, their 'quantum', 'relativistic', and 'classical' behaviour, the entities of space and time, the properties of mass, electric charge, spin and four 'fundamental interactions', now internally unified from the very beginning into one complex-dynamical interaction process between the two protofields. The origin of the well-known persisting 'mysteries' and irreducible separation with 'tangible' reality and between different domains of the canonical fundamental physics (and science in general) is clearly explained as being due to its fundamental reduction of the dynamically multivalued diversity of the real world to the single-valued (unitary), effectively one-dimensional analysis. The practical urgency of transition to the proposed dynamically multivalued description and the ensuing truly adequate knowledge of the universal science of complexity in fundamental physics and beyond is substantiated.

---

*Address for correspondence: Post Box 115, Kiev-30, Ukraine 252030.
 E-mail: kiril@metfiz.freenet.kiev.ua

TABLE OF CONTENTS

Page

# 1. Introduction: Crisis in fundamental physics and the advent of the universal dynamic complexity

Despite their visible diversity and high technical sophistication, the existing schemes of the canonical, basically unitary interpretation of the fundamental physical entities and processes cannot provide an adequate picture of the observed reality. This 'crisis of physics', especially clearly seen in the canonical quantum mechanics, field theory, particle physics, and cosmology, is now readily recognised not only within alternative, so-called 'causal' theories, but also by adherents of the unitary approach itself (e. g. [1-5]).

The modern critical state is a result of development of the fundamental contradictions between the 'classical physics' and the 'new physics' that had started appearing since the end of the nineteenth century. The discovered new types of behaviour, later designated as 'quantum' and 'relativistic' phenomena, had first revealed their absolute incomprehensibility within the notions of classical physics (unitary and regular by definition), but then were somewhat too hastily, superficially, and actually inconsistently 'explained' within the self-proclaimed 'new physics'. Those 'explanations', including the standard 'quantum mechanics' and 'relativity', were based on *formal* correlations between observations and *postulated* mathematical equations, while the detailed qualitative, 'physical' origin of the critically important, properly 'new' effects (like quantum indeterminacy and wave-particle duality or relativistic space-time dependence on motion) was replaced by formally imposed *axioms* about the new 'rules of the game' that may not require any deeper substantiation.

It is important that the 'new physics' dominating since then through all the twentieth century (see e. g. [6]) has always been as fundamentally unitary and regular as the formally 'surpassed' classical physics, which clearly demonstrates that nothing really 'new' was actually introduced.[1] This evidently fictitious 'novelty', in addition attained at the cost of neglecting the demands of elementary consistency and genuine, physical *understanding*, was judiciously and decisively rejected by the actual, original discoverers of the contradictions within the classical physics and the essential elements of the forthcoming new understanding (like Lorentz and Poincaré for 'relativity' and Planck, de Broglie and Schrödinger for 'quantum mechanics'). They clearly saw the problems they discovered themselves and were actively creating the elements of their consistent solution, but could accept only solutions with *at least the same, or larger*, causality (completeness) as that of the old, classical physics, since any true *progress* in science should only *increase* the fundamental understanding of reality. The emerging absolutely new reality certainly demanded construction of a qualitatively new concept providing much *more* intrinsic possibilities for its adequate, *causal* understanding, and therefore its creation needed more time and deeply structured efforts. However, as it often happened within this century of massive 'emancipation' (from any true values) and destructive inclination before any esoteric 'mystification', the appearing 'new wave' of extremely ambitious 'wunderkind' scientists quickly put an end to the 'old-fashioned' doubts of common sense and honesty by presenting them as excesses of conservatism that did not want to give place to the advancing 'progress' of the 'new thinking'. As a result of that kind of progress, we have now many tons of permanently renewed volumes of the most sophisticated constructions of 'mathematical physics' and its 'applications' in 'field theory' and 'quantum mechanics', but as little understanding as ever about the true, physical nature of the elementary building blocks of the universe ('elementary particles'), their dualistic behaviour, 'quantum' and 'wave' character, accompanying 'relativistic effects', and relation to the extremely diverse 'ordinary' world around and within us.

---

[1]Note that at the beginning the term 'new physics' was used rather as a general reference to qualitatively new, non-classical phenomena that were progressively and undeniably emerging in observations as they covered new, previously inaccessible spatio-temporal and energetic scales. This meaning of the term, quite neutral and thus correct as such, was then gradually and tacitly replaced by a much more specific sense referring to particular *interpretations*, and even *type* of interpretation, of those observations which involved the characteristic 'postulated understanding' (or rather *fundamental impossibility* of understanding) based on *intentional* detachment from reality in favour of purely abstract symbols and 'justified' by superficial pretensions for a 'new logic' and the related pseudo-philosophical mysticism.

Moreover, further progress in science can only consist in unification of its various entities and domains, in agreement with the physical unity of the world, and it becomes quite clear now that the deceptively easy road of purely quantitative, 'symbolical' correlations within the 'new physics' cannot bring us to the Temple of the intrinsically unified, truly adequate knowledge, irrespective of the quantity of further efforts. It becomes more and more evident that the same basic deficiency underlies the absence of true understanding at both microscopic (fundamental) and macroscopic levels of reality, and the complete theory of quantum behaviour or gravity can be obtained only within the same universal concept that will provide causal understanding (now also absent) of the involved patterns of 'complex behaviour' in the directly perceived, macroscopic environment.

In this way, the 'advanced symbolism' of the canonical science has brought it to the current deep impasse and accompanying inevitable decay resulting in various forms of technically equipped empiricism and theoretical agnosticism [5]. The issue from that situation can be found only within a qualitatively new concept, or 'paradigm', that should possess the lost universality and causality of the 'classical' knowledge, but now revived at a quite new level of the genuine, *intrinsically complete understanding*. Such a concept has recently appeared in the form of the 'dynamic redundance paradigm' [7], and it was shown that it provides the desired consistent resolution of the fundamental problems of quantum mechanics by extending it to a case of *dynamically complex*, intrinsically *multivalued (redundant)* behaviour and description (see also [8-11]). Moreover, in accord with the above demands, the *same* theory, realising the idea of the really *first-principle approach*, provides the causally complete understanding of the universal gravitation, elementary fields and particles, naturally *emerging* in permanent *dynamic* duality with the attached *material* waves, and a physically transparent interpretation of space, time, mass (energy) and origin of their 'relativistic' relation to motion. The necessary 'new' element is clearly designated as the *unreduced dynamic complexity* in the form of *fundamental dynamic multivaluedness* of well-defined, mutually *incompatible* dynamical regimes/solutions called *realisations* [7,8], and it is important that this *extension* of the old, single-valued concept appears in a formally consistent and physically natural way, causally *completing* the restricted picture of classical physics by many *additional* possibilities of at least the same (and actually much larger) *transparency* and *causality*.

Note that this new concept of complexity, based on the analytically deduced dynamic redundance [7], has nothing to do with the mechanistic, always superficial and contradictory, *imitations* (simulations) of external, empirically registered manifestations (or 'signatures') of complexity within the same, basically single-valued approach of the canonical science that have been appearing in a large number during recent years (see also ref. [12] for a more detailed description of the main species and limitations of that speculative, substitutional 'science of complexity'). In particular, we show that *any* real dynamical system *always* finds itself in a regime of the irreducibly complex behaviour (which is therefore always causally unpredictable to a certain, *sometimes* small, degree), whereas the simulative, unitary 'science of complexity' attributes various externally taken and simplified manifestations of dynamic complexity to highly *specific*, non-universal types of systems defined as 'complex' exclusively by those external special 'signs', so that each of those 'types of complexity' appears to be fundamentally separated from both 'normal', 'non-complex' systems, and other types of 'complex' behaviour (see ref. [7] for more detail). In reality, the imitation of complexity in the canonical, unitary science is practically always reduced to a semi-empirical analysis of computer simulation or (more rarely) experimental results, taking the form of positivistic 'registration' of observations and their mechanistic 'classification', without any more profound understanding of their ultimate, physical origin.

In this article we concentrate on presentation and discussion of the discovered causal, irreducibly *complex-dynamical* origin of the *universal gravitation* and especially the ensuing essential *modification* of values of the well-known *Planckian units* (scales) of space, time, and mass which can have many important theoretical and practical consequences, such as the causally substantiated limitation on expected (mass) *spectrum of elementary particles*. We show that the latter result permits one, in particular, to avoid unnecessary high expenses in accelerator research dedicated to confirmation of many purely speculative assumptions about hypothetical 'new particles' with excessively high masses. This is a clear demonstration of the practical utility of the proposed unified and causally complete theory.

We start by recalling, in section 2, the main lines of our causal version of the world construction providing the natural source of universal gravitation and its intrinsically *quantum* (= dynamically discrete) character. It is shown that if we consider an initially *homogeneous* system of two *material* 'protofields' *uniformly* attracted to each other as the simplest possible construction of the world, then it can already provide the hierarchy of self-sustained, and highly *inhomogeneous*, complex-dynamical processes appearing as the observed particles and bodies. The lowest level of this 'universal hierarchy of complexity' is formed by the 'elementary particles' each of them causally emerging as the process of 'quantum beat' in the system of two homogeneously coupled protofields. The *quantum beat* can be described as an unceasing series of cycles of physically real 'reduction' (self-amplified, dynamical auto-squeeze) of an extended portion of the protofields to a small volume around a centre of reduction, forming the physical 'space point', and simultaneously the 'corpuscular' state of the thus emerging 'field-particle', and followed by the reverse extension. Each centre of reduction is 'chosen' by the system in a *causally random* fashion among *many* other equally *probable* centres, which provides the purely *dynamical* basis for quantum *indeterminacy*. The consecutive cycles of reduction-extension form the elementary physical *events* marking the *discrete* intervals of the causally emerging, naturally and *unceasingly* 'flowing' *time* which is *irreversible* by definition (due to the causally unpredictable choice of each next centre of reduction). The unceasing *complex-dynamical process* of quantum beat constitutes the physical essence of every (massive) *elementary field-particle*, providing the natural and consistent explanation for the most 'mysterious' observed properties of 'quantum' objects, those of *wave-particle duality* and *quantization* (discreteness), since the field-particle *permanently changes its state* from the squeezed, 'corpuscular' state to the extended 'wave' and back, and such *global* transformation can happen only in a *dynamically* discrete, or 'quantum' fashion. It is important that quantum beat processes within different elementary particles of the world are synchronised in phase (up to inversion of phase), which explains the universality of both causal time and emerging property of electric charge with two its 'opposite' species [7].

The key process of autonomous, catastrophic auto-squeeze (and the reverse extension) in an a priori absolutely homogeneous, simplest possible system of interacting primal entities can be obtained due to a *universally* nonperturbative analysis within the 'generalised effective potential method' [7,8,13], where we do not perform the usual [14] unjustified rejection of the self-consistent dependence of the effective potential on the solutions to be found which just provides the intrinsic *dynamic instability* and universally defined *essential nonlinearity* of a *generic* system with interaction through the feedback loops of the *unreduced* interaction process. The same mechanism determines the related fundamental property of *dynamic redundance* (multivaluedness) and the ensuing causal randomness, since the initial *homogeneous* system of interacting protofields provides *many* potential centres of reduction, each of them being equally probable for any individual event of reduction.

It is clear that each individual quantum beat process appearing as the elementary field-particle will change the local 'mechanical' ('elastic') properties of the protofields and thus influence the characteristics of other close enough quantum beat processes (and first of all, their *temporal rate* determining the property of *mass-energy* [7,11]), which appears as long-range types of 'fundamental' *interaction(s)*, or 'forces', between particles. The two original protofields form the two basic 'sides', or 'sheets', of the *physically unified* world structure that can be called 'world sandwich', and the *directly* perceivable reality is situated at only one side of the 'sandwich' referred to as the *electromagnetic* (e/m) protofield (because its perturbations appear as various e/m effects, e. g. waves, or 'photons'). The *e/m interactions* are transmitted, by the mentioned mechanism, through the e/m side of the couple in the form of 'exchange of photons', now being causally specified [7]. The other side of the 'sandwich' is formed by a medium/protofield of a different physical nature designated as 'gravitational'. The gravitational protofield cannot be directly perceived within this world, but it is an equally real partner in the complex-dynamical reduction process and therefore it provides another, universal, way of interaction 'transmission' between any particles with non-zero mass-energy realised, within the same *causal* mechanism of 'action at a distance', by changing local mechanical properties of the gravitational medium. This interaction by *complex-dynamical* intermediation through the 'hidden', but *material* degrees of freedom of the gravitational protofield is observed as the *universal gravitation* (attraction between *any* massive elementary particles and bodies constituted from them).

The proposed mechanism of gravitation can be considered as the causally complete extension and essential modification of the standard interpretation of the 'general' relativity within *purely abstract* 'geometry' of abstract 'manifold' of ambiguously *mixed* space and time, and therefore contrary to the latter, it *naturally* provides *any* gravitational effects with internal 'quantum' (complex-dynamical) structure *directly* perceivable only at the ultimately small scales of space and time. After specifying these and other details of the proposed causal understanding of gravity, we reveal, in section 3, its consequences for the causal interpretation of the *Planckian units* which involves also essential *modification* (renormalisation) of the values of those units of space, time, and mass (those readers who are more interested in this particular consequence of the general picture can skip the review of section 2 and pass directly to section 3). It is clear that this result will induce many important changes in multiple considerations which use Planckian units. One of the most significant, practically important and timely conclusions concerns the mass spectrum of elementary particles existing in nature. Since we show that the properly renormalised Planckian unit of mass, the 'Planck mass', is much smaller than the conventional value and is just close, by order of magnitude, to the heaviest observed particles (with the mass of the order of 100 GeV, in energy units), it follows that one should not try to build new, monstrous and expensive, 'super-accelerators' in the search for particles with much larger masses predicted by various 'heuristic', purely abstract theories of the type 'what if'.

In general, by clearly specifying the basic *physical nature* of the elementary particle as such, the holistic approach of the universal concept of complexity actually provides the necessary unifying 'rule', or 'ordering', for the fundamental levels of reality which permits one to know *objectively* what and why exactly can *physically* exist in nature (which means that the unreduced concept of complexity provides an objective and universal *criterion of truth* [7] applied here for the lowest levels of reality, but unifying actually the whole diversity of being). The complex-dynamical 'order' is both consistently substantiated and possesses the necessary pronounced 'asymmetry' widely present in the observed entities, but only artificially, mechanically inserted into canonical, basically single-valued (unitary) theories e. g. in the form of 'spontaneously broken symmetry'. There is no wonder that the unifying 'order of the universe' is obtained in our approach together with the 'unified interaction' helplessly sought for by the scholar science in a casual choice performed among ugly mechanistic constructions of the 'mathematical physics' irreducibly detached from reality. The unique, 'driving' (i. e. structure-*creating*) interaction between the two protofields *naturally and consistently* realises the causal extension of any 'Grand

Unification' of the unitary science by giving rise to the four observed 'fundamental' types of interaction between particles emerging as different, and well-defined, *dynamical modes* (or, again, 'realisations') of the single *complex-dynamical process* of the protofield interaction (the two short-range modes within the e/m and gravitational protofields, the 'weak' and 'strong' interactions respectively, naturally complement the two corresponding long-range modes, the e/m and gravitational interactions, see the end of section 2.1).

Equipped with these results, we return then, in section 4, to discussion of the emerging extended causality and universality of the new knowledge actually continuing the tradition of the 'classical science', so disgracefully and ruthlessly interrupted, at the beginning of the now ending century, by the superficial symbolism of the 'new physics'.

## 2. Unified causal origin of universal gravitation, elementary particles, their quantum behaviour, and relativity of space and time

### 2.1. The world as a system of two interacting protofields and the complex-dynamical process of quantum beat

Any really causal, first-principle description of the world should start with the well-defined absolute minimum of physically real initial entities and show, by using the elementary logical rules representing those entities, how the observed diversity of structures naturally, autonomously emerges from that initial configuration, without intervention of any external 'influences' or additional, artificially imposed 'principles' of higher order (such as 'principle of least action', other cases of 'variational principle', 'Born's probability rule', or 'principle of relativity' in the canonical science). The necessary 'non-deducible postulate' of such really first-principle theory, realising the causal extension of the 'Gödel incompleteness conjecture' [7], is provided by the *same* assumption about this 'minimal initial construction' lying in the basis of the World and actually determining its physical 'type' and material 'quality' among other conceivable worlds in the continuous hierarchy of creation. Note also that such description will automatically possess a 'cosmological' character revealing *simultaneously* the nature of an existing structural element (object) and the actual *history* of its appearance: in the resulting 'universal science of complexity' [7] the properties of an object cannot be separated from the process of its emergence because *any* entity with all its properties is explicitly *obtained* as a result of a well-specified development of the hierarchy of interaction processes of initial (primal) entities. This *emergent* character of the new knowledge is qualitatively superior with respect to that of the canonical science inevitably separated into a large number of effectively linear (i. e. single-valued and therefore one-dimensional) stages/entities/domains that are first positivistically 'guessed', in the form of simplified, purely abstract constructions, and then should be mechanically 'put together' by many additional, artificially imposed, and therefore 'mysterious' postulates (this program actually fails even within its own restricted logic and starting already from the lowest levels of fundamental entities).

We argue that the absolutely necessary minimum of entities for the initial configuration is provided by two physically real ('material'), *a priory* homogeneous distributed media, or 'protofields', endowed with an attractive interaction between them. Indeed, it should be clear that any number of *non*-interacting entities cannot provide autonomous emergence of anything new (this statement is additionally confirmed by further analysis). From the other hand, only one 'self-interacting' field is always a manifestation of a basic interaction of two more fundamental entities (this will also be clearly seen from the results below). The demand for absolute initial *uniformity* of the protofields and their interaction is also necessary for a first-principle description, where any structures should be explicitly *obtained* within natural development of the interaction process. For this reason, acceptance of any 'particles' as initial entities, performed in a number of approaches (such as 'Bohmian mechanics' [15-18]), seems to be less consistent, since it does not liberate us from the necessary assumption about other physical entities (eventually always reduced to 'fields') which should constitute the 'material' and 'environment' of the particles and thus belong to the *same* 'level' of reality. Our initial protofields, being really existing, physically 'tangible' entities should also possess some internal structure, but contrary to the situation with particles it may (and in fact necessarily should) belong to some *lower* (more 'fundamental') level of reality, sufficiently well separated from the results of protofield interaction in order to remain unresolvable from 'inside' of this world. This separation is possible due to the fundamental property of dynamical *discreteness* of any emerging structures and their 'levels of complexity', which is self-consistently confirmed in further analysis of the (generic) interaction process. Note also that the initial homogeneous, 'free' state of the protofields before the beginning of interaction process can be consistently defined as the physically real, but unobservable *ether*: the protofields/media can become 'observable' only after their inhomogeneous perturbations (e. g. e/m waves or 'particles') emerge as a result of interaction, since any 'observation' is realised eventually as interaction between some of those 'perturbations'.

The minimal initial configuration of interacting protofields can be expressed by the *existence equation* which does not add any new assumption about the system, but simply fixes the fact of its existence as a *single*, internally unified ('holistic') system that cannot be reduced to the trivial sum of independent components (protofields in this case):

$$[h_e(q) + h_g(\xi) + V_{eg}(q,\xi)]\Psi(q,\xi) = E\Psi(q,\xi) , \qquad (1)$$

where the *state-function* $\Psi(q,\xi)$ just characterises the *compound* state of this *integrated* system of interacting primal entities $q$ and $\xi$ (it can reflect the field magnitude of their emerging physical mixture specified below), $q$ is the electromagnetic and $\xi$ the gravitational degrees of freedom, $h_e(q)$ and $h_g(\xi)$ are the generalised Hamiltonians (or other actually measured functions) for the 'free' (non-interacting) media, $V_{eg}(q,\xi)$ is a generally unspecified (though eventually attractive and binding) interaction potential between the fields of $q$ and $\xi$, and $E$ is the energy (or other property corresponding to the selected 'generalised Hamiltonian' and representing a measure of dynamic complexity fixed at its lower level, see further analysis). We emphasize once more that the existence equation, eq. (1), has a quite general, almost 'symbolical' character and states simply that the considered system with interaction 'exists as a (nonseparable) whole'. In particular, the familiar form of eq. (1) is chosen for convenience and could be replaced by another symbolical representation with the same meaning, so that, for example, the signs "+" in this equation need not be necessarily interpreted directly, in the sense of simple additivity of the participating entities.[2] This means also that this equation and the results of its analysis are applicable to *any* system of *arbitrary* interacting entities/objects, and such uniformity is important for understanding of the self-consistent, quasi-autonomous further development of the holistic world structure in the form of the 'universal hierarchy (arborescence) of dynamic complexity' [7] which avoids, in particular, any 'bizarre' ruptures between 'quantum' and 'classical' levels of being (we specify this statement later).

Now we can show, using a *universally* nonperturbative method of analysis, that such quite generally described interaction process results in a well-structured sequence of *catastrophic* internal transformations of the system constituting its *complex dynamics* and determined by unceasing reappearance of intrinsic *dynamic instability* in a system with interaction. In order to specify the structure of the protofield entanglement within the single object, we begin by expanding the system state-function over a complete system of eigenfunctions, $\{\phi_n(q)\}$, for the free e/m protofield:

$$\Psi(q,\xi) = \sum_n \psi_n(\xi)\phi_n(q) , \tag{2a}$$

$$h_e \phi_n(q) = \varepsilon_n \phi_n(q) . \tag{2b}$$

The eigen-solutions $\{\phi_n(q)\}$, $\{\varepsilon_n\}$ represent 'possible' modes of the protofield existence corresponding to the chosen 'measured quantity' $h_e$. This choice is not unique, and formally any complete system of functions would be acceptable, but for a particular system with interaction it should rather correspond to the structures that are expected to really emerge in course of interaction, where the system performs a natural 'measurement on itself'. In our case, we can expect the emergence of localised spatial 'points' and the corresponding 'corpuscular' states (which will be confirmed below), and therefore can take $\phi_n(q)$ in the form of localised, $\delta$-like functions, $\phi_n(q) \propto \delta(q - q_n)$, marking the future 'coordinates' $\varepsilon_n = q_n$. Note that this (or another) particular choice does not involve any additional postulate about the system: for the localised $\phi_n(q)$, the expansion of eqs. (2) corresponds to presentation of a 'distributed medium' (field) in the form of a sum of its small localised pieces (this *formal* discretization will simply help us to reveal the real, dynamical *quantization* of the system into well-defined local structures). For the moment we shall continue the analysis in the general form, evoking the particular properties of $\phi_n(q)$ when it will actually become important.

---

[2]Such absolutely general relation between the interacting entities, $q$ and $\xi$, expressing the fact of their 'nonseparable' interaction, can be presented simply as $\{q,\xi\}$, implying that $\{q,\xi\} \neq \{q\},\{\xi\}$ at any scale. However, it can be shown [7] that any interaction process can be presented in a Hamiltonian form, obtained as the extended version of the Hamilton-Jacobi equation, where the (generalised) Hamiltonian characterises the temporal rate of realisation change within the universal 'complexity unfolding' process. In any case, it will be sufficient, at this stage, to consider eq. (1) as adequately reflecting the assumed 'minimal' configuration and avoiding any additional restrictions.

Insertion of expression (2a) for $\Psi(q,\xi)$ into eq. (1) transforms the existence equation into a system of equations for $\psi_n(\xi)$ [7-10] having generally the same meaning, but expressed now as 'many-body' interaction between different 'pieces' (or 'existence modes') of both protofields.[3] We now analyse this interaction process by first selecting one equation of the system that describes the behaviour of certain, say, 'zeroth' state-function component, $\psi_0(\xi)$, in its interaction with other components. We then express these other components through $\psi_0(\xi)$ from 'their' equations, using the standard Green function technique [7-10,13,14], and insert the results into the first equation, which transforms it into a formally 'closed' form:

$$[h_g(\xi) + V_{\text{eff}}(\xi;\eta)]\psi_0(\xi) = \eta\psi_0(\xi), \qquad (3)$$

where $\eta \equiv E - \varepsilon_0$ ($\eta_n \equiv E - \varepsilon_n$ for $n \neq 0$), the *effective (interaction) potential* (EP), $V_{\text{eff}}(\xi;\eta)$, is given by

$$V_{\text{eff}}(\xi;\eta) = V_{00}(\xi) + \hat{V}(\xi;\eta), \quad \hat{V}(\xi;\eta)\psi_0(\xi) = \int_{\Omega_\xi} d\xi' V(\xi,\xi';\eta)\psi_0(\xi'), \qquad (4a)$$

$$V(\xi,\xi';\eta) \equiv \sum_{n,i} \frac{V_{0n}(\xi)\psi_{ni}^0(\xi)V_{n0}(\xi')\psi_{ni}^{0*}(\xi')}{\eta - \eta_{ni}^0 - \varepsilon_{n0}}, \quad \varepsilon_{n0} \equiv \varepsilon_n - \varepsilon_0, \qquad (4b)$$

$$V_{nn'}(\xi) \equiv \int_{\Omega_q} dq\, \phi_n^*(q)V_{eg}(q,\xi)\phi_{n'}(q), \qquad (4c)$$

and $\{\psi_{ni}^0(\xi)\}$, $\{\eta_{ni}^0\}$ are the complete sets of eigenfunctions and eigenvalues for an 'auxiliary', truncated system of equations (where $n \neq 0$)

$$[h_g(\xi) + V_{nn}(\xi)]\psi_n(\xi) + \sum_{n' \neq n} V_{nn'}(\xi)\psi_{n'}(\xi) = \eta_n\psi_n(\xi). \qquad (5)$$

---

[3]Note that this transformation can be performed not only for linearly acting quantities $h_e$, but they should satisfy the 'principle of superposition', so that for example $h_e[\sum_n \psi_n(\xi)\phi_n(q)] = \sum_n \psi_n(\xi)h_e[\phi_n(q)]$. This condition corresponds to the natural demand for the elementary 'eigen-modes' of the 'free' primal entity to be 'independent' (not entangling with each other) because otherwise the complex-dynamical, chaotic behaviour will result already at the level of internal dynamics of this 'free' field, and one would need to 'descend' to a still lower level of reality in search of its understanding, etc. An example of 'nonlinear' superposable modes is provided by the canonical solitons, even though we need not actually imply anything of this kind for the considered case of interacting protofields (the 'free' e/m protofield could well be 'linear', at the observable scale of a problem). In any case, the transition from one existence equation to the equivalent system of equations is rather a matter of technical convenience reduced to passing from variable $q$ to the 'mode number' $n$, without assuming its physically discrete character (the real interaction process discreteness will appear below in a different, purely dynamical way).

Note that the generalised 'method of substitution of variables' of this kind, where the problem is formally reduced to an equation for only one 'degree of freedom', but the other 'degrees' actually enter into the expression for the effective potential in that equation, is well known in the theory of scattering and many-body problem in general under the name of "optical (effective) potential method" [14]. The general solution of a problem is obtained by insertion of the found eigen-solutions of the 'effective' eq. (3), $\{\psi_{0i}(\xi)\}$, $\{\eta_i\}$ into expressions for $\psi_n(\xi)$ ($n \neq 0$) mentioned above and of all of them into expression (2a) for $\Psi(q,\xi)$ [7-10]. It is clear that the found 'solution' is not really closed, since it depends itself on unknown solutions. Therefore in the standard optical potential method, the above expressions for EP are then always explicitly simplified by using one or another version of perturbation theory [14]. These procedures involve actual elimination of the 'inconvenient' EP dependence on the eigen-solutions to be found and dynamic entanglement between different protofield eigen-modes $\psi_{ni}^0(\xi)$, $\phi_n(q)$ (see eq. (4b)), determining the problem 'nonseparability'. However, it is not difficult to see that it is precisely that self-consistent EP dependence on the effective equation solutions determined, in their turn, by EP which produces the qualitatively new, generic, and intrinsically complete effects of (unreduced) *dynamic complexity*. The key feature is the phenomenon of *dynamic redundance* (or *multivaluedness*) which appears due to the EP dependence on the eigenvalues to be found ($\eta$ in eq. (4b)) and consists in existence of *many* solutions of eq. (3), each of them being 'complete' in the ordinary sense and therefore *incompatible* with other solutions. Therefore we call each of such exhaustively defined solutions *realisation* of the system. The formal reason for such 'solution multiplication' (or system 'splitting' into multiple realisations) is the evidently increased highest eigenvalue power with which it enters the 'characteristic equation' for the eigenvalue problem for eq. (3), which can also be confirmed by a graphical method of solution [7-10,13]. We shall enumerate the quantities belonging to different realisations by an additional index, *r*, which originally appears in the eigen-solutions, $\{\psi_{0i}^r(\xi)\}$, $\{\eta_i^r\}$, of the effective equation, eq. (3), because of the described dynamical multiplication process. It is *this* extended set of solution-realisations that appears to be 'normal', generic, and really complete, rather than the ordinary, highly reduced set of the canonical perturbational solutions corresponding to *only one* (and distorted) realisation, while all other, equally real solution-realisations are groundlessly omitted in any canonical approach (this explains, in particular, the well-known difficulties with convergence of perturbative expansions, including the apparently 'good' ones).

The genuine system realisations, obtained from the unreduced effective equation, differ from each other in their characteristic parameter values (such as the EP barrier amplitude and width) and the relative differences are generally *not* small. Being incompatible, but *equally* valid (and thus probable), system realisations should *permanently replace* each other, in a fundamentally *unpredictable* sequence (order and moments of appearance), which provides the naturally emerging *causal origin of randomness* in the world and the ensuing definitions of (true) *dynamical chaos* and *complexity* [7-10,13]. It is important that, due to the real basis of the dynamic redundance phenomenon, those fundamental notions can be defined in a physically transparent fashion, contrary to their mechanistic *imitations* within the single-valued

'science of complexity' using infinitely adjustable, but only external sophistication of abstract, *purely* mathematical constructions. Thus, the causal dynamic randomness will evidently appear in observable system behaviour as an unceasing sequence of irregular changes of dynamic regimes (or their measured parameters), corresponding to incompatible realisations, while the related causal dynamic complexity is determined by a growing function of the number of realisations (or the rate of their change) and characterises the diversity of possible (and *really* appearing) versions of system behaviour. Complexity is equal to zero for *any* single-valued, unitary, regular (i. e. totally predictable) dynamics, which includes *all* the imitations of the canonical science, but actually can never occur in nature in exact, pure form (and therefore even the simplest real phenomena, like independent elementary particles and their gravitational interaction, are represented by processes with non-zero complexity and irreducible internal unpredictability).

A closely related aspect of the dynamic complexity thus defined involves the 'self-organised' internal structure (configuration) of each realisation. It appears as a densely interwoven, or *entangled*, basically irregular mixture of the interacting degrees of freedom (protofields in our case), formed in a purely *dynamic*, autonomous fashion in course of the unreduced interaction process and *concentrated* around certain, *randomly* 'chosen' centre representing a physical 'point' of the naturally emerging *space* (of this level of complexity). This *physically real* concentration, or squeeze, or *reduction* of the interacting protofields occupying a particular, $r$-th realisation (in the process of their change) is described by expression for the wave function, $\Psi_r(q,\xi)$, and measured density, $\rho_r(q,\xi)$, of that realisation obtained from eq. (2a) by the above EP formalism [7,10]:

$$\rho_r(q,\xi) = |\Psi_r(q,\xi)|^2 , \quad \Psi_r(q,\xi) = \qquad (6)$$

$$= \sum_i c_i^r \left[ \phi_0(q)\psi_{0i}^r(\xi) + \sum_{n,i'} \frac{\phi_n(q)\psi_{ni'}^0(\xi) \int d\xi' \psi_{ni'}^{0*}(\xi')V_{n0}(\xi')\psi_{0i}^r(\xi')}{\eta_i^r - \eta_{ni'}^0 - \varepsilon_{n0}} \right],$$

where $c_i^r$ are numerical coefficients that should be determined from the 'dynamical' boundary conditions [7,10]. It is easy to see from this expression that the compound field magnitude for each realisation, represented by $\Psi_r(q,\xi)$, is *physically* concentrated around certain 'preferred' eigenvalue (for that particular realisation), while gradually decreasing with the distance from it, measured in the space formed by eigenvalues [7,9,10]. If we choose the basis of initial expansion in the form of localised functions mentioned above (i. e. $\phi_n(q) \propto \delta(q - q_n)$ and $\varepsilon_n = q_n$), then we see that due to the resonant denominators and 'cutting' matrix elements in the numerators of eq. (6) only a small number of terms with close *n* can have appreciable magnitude for a fixed *r*, which corresponds to the dynamical squeeze towards a really emerging, non-formal, physically 'produced' point, also representing the *corpuscular*, particle-like state of the system. It is important that the latter is obtained as a result of *dynamic entanglement* (or physical 'mixture') of the interacting degrees of freedom, $q$ and $\xi$ (described by the entangled products of the corresponding functions in eq. (6)), which provides the emerging new structure (physical point, or *field-particle*) with the necessary *new quality* not reduced to simple addition of the qualities of the interaction partners (protofields). This physical, dynamic entanglement of the interacting entities continues in a hierarchy of ever smaller (unobservable) scales forming a *fractal* internal configuration of the emerging elementary field-particle (see below). The preferred eigenvalue for the *r*-th realisation is given approximately by those members of the found eigenvalue set $\{\eta_i^r\}$ for which the denominators and numerators in eq. (6) have generally close values further adjusted through similar eigenvalue dependence of EP, eqs. (4). It is important that the latter has the same form as the above state-function expression:

$$V_{\text{eff}}(\xi;\eta_i^r)\psi_{0i}^r(\xi) = V_{00}(\xi)\psi_{0i}^r(\xi) + \sum_{n,i'} \frac{V_{0n}(\xi)\psi_{ni'}^0(\xi) \int d\xi' \psi_{ni'}^{0*}(\xi')V_{n0}(\xi')\psi_{0i}^r(\xi')}{\eta_i^r - \eta_{ni'}^0 - \varepsilon_{n0}} . \quad (7)$$

This means that the effective interaction for the *r*-th realisation has the largest amplitude, i. e. dynamically produces a potential well, just around the same eigenvalue for which the state-function for that realisation has its 'centre of concentration', which has the return influence on the state-function reduction in eq. (6). This self-consistent eigen-solution formation adequately reproduces the self-amplifying, catastrophic reduction process, thus clarifying the physical meaning and demonstrating the reality of the 'effective' description of a problem, where a 'potential well' is *dynamically formed* within the *same* nonlinear reduction process as the states 'trapped' in it.

    Due to the unified character of the description used, this result shows that a generic (nontrivial) interaction process involves the natural appearance of a 'standard' *dynamic instability* leading to the catastrophic reduction-entanglement. The instability results from the self-developing hierarchy of interaction feedback loops properly taken into account in the universally nonperturbative analysis of the generalised EP method (by the above self-consistent formation of eigen-solutions) and can also be considered as *unified* definition of the *essential*, objectively interpreted *nonlinearity*, as opposed to its various superficial definitions in the canonical, single-valued, perturbational science. Such universal dynamic nonlinearity implies thus a 'self-interaction' process that can be externally 'hidden', but always actually develops in any generic, non-trivial system with interaction and leads to formation of many incompatible, and therefore permanently replacing each other, structure-realisations constituted from a densely entangled, fractal mixture of interacting entities (degrees

of freedom). This dynamically structured, internally unstable, self-developing nonlinearity is quite different from the mechanistically fixed canonical 'nonlinearity' implying just a formal 'curvature' of an external form, particular algebraic type of function, or an isolated 'sign' (like the broken 'superposition principle'), which properties are readily attributed to many single-valued, 'exact' solutions of the unitary science (like solitons): from the point of view of our unreduced dynamic complexity all of them are fundamentally not more nonlinear, than a straight line.

For the particular considered system of interacting protofields, the above universal mechanism of dynamic instability and nonlinearity can be presented in an especially transparent form operating exclusively with physically evident, irreducible properties. In order to describe it, we just imagine that the e/m protofield, *uniformly* attracted to the gravitational protofield, produces a very small (in principle, infinitesimal) fluctuation corresponding to a slight increase of its density around one particular location. Then the elements of the gravitational protofield situated around that locality will be *additionally* attracted by the e/m protofield fluctuation and therefore also produce a relative increase of their concentration with respect to the surrounding homogeneous value. But this will provoke further reaction of the same type within the e/m protofield, that is its initial fluctuation will grow because of additional attraction of the increased gravitational protofield concentration. The process will evidently continue in a self-amplified fashion until the opposite forces of internal pressure compensate the forces of attraction. At that 'moment of equilibrium' we obtain the corpuscular, particle-like state of the system in the form of a highly concentrated, localised mixture of the interacting protofields (though perceived by us only from its 'e/m side') which forms also the elementary physical 'point of space' emerging thus *simultaneously* with the particle-object (cf. the Cartesian concept of matter reduced to its spatial extension [7]). Since the initial (infinitesimal) fluctuation can happen with equal probability at any location (future space point), we obtain the natural source of randomness at this fundamental level of being, referred to as 'dynamic redundance'. It is clear that the number of such incompatible possibilities/realisations, $N_\Re$, determining the 'degree of freedom of choice' (or 'causal randomness') of a system, is equal to the effective number of possible 'points' participating in the interaction process (defined formally as the number, $N_g$, of the gravitational protofield eigen-modes $\psi_{ni}^0(\xi)$ effectively contributing to expansions of eqs. (4b),(6),(7), so that $N_\Re = N_g$). This total realisation number forms a quantitative measure of dynamic complexity at its given level and is a manifestation of the 'magnitude of (unreduced) interaction'; in the case of electro-gravitational coupling within the electron the number of realisations is determined [7] by the reciprocal 'fine structure constant' ($\alpha \approx 1/137$), $N_\Re \sim 2\pi/\alpha$.

Note that the described physically evident and quite simple origin of the fundamental dynamic multivaluedness, the ensuing instability and randomness inherent in any interaction process clearly demonstrate the striking insufficiency (not to say a gross blunder!) of the whole canonical, unitary science completely ignoring, within its irreducibly perturbational approach, all the system realisations but one, thus reducing any system behaviour, and therefore also the whole world, to a mechanical sum of separable, effectively *one-dimensional* dynamical regimes. This evident deficiency cannot be eliminated within the existing schemes of the so-called 'science of complexity' (including various versions of 'catastrophes', 'bifurcations', and 'attractors') which only try to *imitate* the true multivaluedness and its consequences within the same fundamentally *single-valued*, and thus unitary approach, almost invariably used in a semi-empirical, computer-simulation form (see [7] for more detail).

As we mentioned above, the unified mechanism of the dynamic nonlinearity development necessary leads to emergence of a hierarchical, *fractal structure* of entanglement (mixture) of the interacting entities within each formed realisation of the new object (elementary field-particle in our case). The details of this process are described by the corresponding further development of the same EP formalism. Namely, each of the obtained realisations of the first (basic) level will undergo dynamical splitting into incompatible (and thus also probabilistically changing) realisations of a deeper (more fine-grained) level due to the (implicit) dependence of the first-level solutions and the solutions of the auxiliary system of equations, eqs. (5), on the eigen-solutions to be found. This is actually the same mechanism that leads to the first level of splitting, but we could neglect this more fine-grained structure, while analysing the emergence of the more 'visible' first-level realisations (thus, this finer structures are not actually resolved for the elementary field-particles constituting, however, the necessary 'filling' of their internal structure). This process can be compared to investigation of the structure of a tree when one analyses all the branches at a given level, while gradually moving towards finer levels, which permits one to take into account *all* the tree branches. Contrary to this, the canonical, single-valued (and therefore inevitably perturbative and effectively one-dimensional) approach can treat *only one* branch-realisation at each level of the hierarchy, while the very existence of *many* levels of the hierarchy is as inexplicable as 'simultaneous' existence of *many* branches at *each* level and can be accepted only as an *empirical* fact. The essential difference of our dynamical fractal from the 'fractals' semi-empirically introduced in the unitary science ([19], see also [7] for more details) is its *causally probabilistic* character and totally *dynamical* ('spontaneous') origin of any level of its structure, which provides a natural source of its development both 'in deep' (to lower levels of complexity forming the fine-grained 'internal structure' of the object), and 'upwards' (to higher levels of complexity giving the course-grained hierarchy of new objects [7]). Therefore the fundamental dynamical fractal of a problem, as it emerges within the dynamic redundance paradigm, is an unpredictably changing, intrinsically self-developing, and therefore *complete*, *solution* of the corresponding dynamic equation which is explicitly (analytically) *obtained* within the unreduced, nonperturbative and universal EP method, without any external, artificial introduction of probabilities, time, space, etc. (instead, all those notions naturally emerge together with the fractal structure, see also below). Note that the mentioned completeness of the fundamental dynamical fractal reflects, in a 'geometrical' formulation, the intrinsic *completeness of realisation set* obtained within the FMDF paradigm at any given 'level of complexity' (or 'level of fractal structure'), which means that the found multivalued solutions provide the really adequate description of reality.

This important property results from the fact that realisation sets at different levels of fractal branching (EP splitting) are obtained as the natural development of the unreduced interaction process, where *all* the details are taken into account. One can only omit, within this approach, an interaction with yet another entity (degree of freedom), but for the given interaction partners we explicitly obtain all the existing possibilities/realisations, which is confirmed by the ultimately great, qualitatively exhaustive number of versions of behaviour provided by the dynamically probabilistic fractal structure described above (one simply does not need to have more of them to cover any conceivable type of behaviour). The property of completeness can also be considered as a manifestation of the *complexity conservation law* [7] stating that the interacting modes can only change their form while producing the realisations of the new level of complexity, but not their total number, so that the number of realisations, $N_\Re$, cannot be different *in principle* from the number of interacting eigen-modes, $N_g$. The conservation of complexity constitutes that universal non-mechanistic 'order' (or 'symmetry') which is necessary for existence of a viable, meaningful world: any really 'spontaneous' emergence of entities 'from nothing' could correspond only to an ultimately irregular world dynamics with practical absence of any structure.

The dynamically emerging fractal structure in a system with interaction is also important as a 'catalyst' of realisation change process. The basic origin of permanent system 'switch' between different realisations is the dynamic redundance itself, but the particular physical transition of the system from one realisation to another is largely facilitated due to the fractal 'network' of the entangled interaction partners surrounding each realisation. Its existence excludes any 'absolute' stability of a system realisation that should inevitably be destroyed due eventually to the same *unified nonlinearity* which has led to its formation. In the case of interacting protofields, the phase of maximum contraction ('corpuscular' state of the field-particle) will be immediately followed by the reverse phase of *extension* to the quasi-free state (close to the initial one) accompanied by the transient *disentanglement* of the interacting protofields. The extension is driven by the never stopping attraction of each of the entangled protofields to the neighbouring parts of its interaction partner which is realised and gradually amplified precisely through the fine fractal network of the protofields around the localised state-realisation, while the disentanglement is the inevitable way of physical development of this process. As a result of extension, the system of interacting protofields finds itself in a particular state, also highly unstable (and highly irregular), that can be considered as a special system realisation playing actually the role of an 'intermediate' state through which the system performs its chaotic transitions (or 'jumps') between the consecutive 'regular' (localised) realisations. Indeed, it is clear that this extended state of the interacting protofields will again collapse to a new dynamically squeezed state, and the cycle of realisation change will infinitely repeat itself. We also call the intermediate realisation 'main realisation', since it can be considered as the prototype for the (hypothetical) single realisation in the (unrealistic) unitary description [7], and actually it plays the role of a complex-dynamical 'glue' that 'bonds together' in one process all the (otherwise independent) 'regular' realisations of the system. All these details are described by the same equations of the unreduced EP formalism (see [7-9]) actually emerging as a specific, delocalised configuration of the eigenstates contributing to the main realisation. Note that the fine dynamical 'connection' between the individual 'regular' realisations demonstrates once more the fundamental difference between the causally complete presentation of the multivalued reality and any canonical, single-valued description within a perturbative 'expansion', as well as the universal origin of the well known 'divergence' of the latter. In reality, any nontrivial (i. e. non-one-dimensional) system with interaction is permanently 'split' into its many 'versions' (realisations), in *every* point of its 'parameter space', and any perturbational 'approximation' is qualitatively *wrong*, since it *always* ignores all the versions but one by ruthlessly cutting the dynamical connections actually contained in the equations of motion. By 'self-consistently' justifying their validity (through the smallness of the immediate corrections, etc.), the perturbational approaches of the canonical science inevitably get into an elementary logical trap of the 'vicious circle' type, which becomes evident when a more thorough investigation shows that the perturbational expansion as a whole invariably diverges (because of the self-amplified influence of the rejected 'small' connection terms, as it becomes clear within the unreduced analysis).

The dynamically driven squeeze-entanglement of the interacting protofields provides also a mechanism of the natural appearance of a curling motion *within* the field-particle known from observations as *spin*. Without going here into much detail (see [7,10]), we note only the fundamental *causal* origin of spin within the dynamic redundance paradigm and its direct relation to the unified dynamic nonlinearity of the protofield interaction process. The rotation of the squeezed and entangled e/m protofield can be attributed to its shear instability amplified by the internal fractal entanglement with gravitational protofield. This mechanism is generally similar to hydrodynamical vortex formation when a liquid is forced to pass by a small hole, but in our case we have a much more 'nonlinear', catastrophically growing phenomenon. The unique role of the dynamic instability/nonlinearity of the unreduced FMDF interaction process is that it *dynamically* creates an *infinite* sequence of those 'small holes' starting from an *extended* configuration of the *physically real* protofields, whereas the canonical perturbative analysis would give only one, more or less homogeneous 'fall' of the protofields one onto another. This mechanism emphasises the indispensable role of the described *unreduced* interaction *nonlinearity* in the causal origin of such a 'mysterious' and 'quantum' property as spin. That is why any of the existing attempts to describe spin within a unitary, perturbational approach cannot succeed in its consistent introduction, and they are obliged to mechanically *postulate* the property of a 'localised torsion' in each point of a field; however, the canonical *dimensionless* 'point' cannot 'rotate' (one *should* have at least a prototype of particular *matter* that forms the rotational motion), even though it is also evident that the rotation of spin should be concentrated in a vanishingly small domain, which devaluates all the 'causal' unitary theories with 'rotating balls' and shows why the single-valued science cannot consistently describe this explicit manifestation of the unreduced dynamic complexity in principle. We have here just another manifestation of the unitary science difficulty with the wave-particle duality, which is not surprising taking into account that spin and reduction-extension oscillation form two inseparably related aspects of the same dynamic complexity of the protofield interaction process (which is reflected in the fact that the elementary quanta of both action and angular momentum of spin are determined by the same, universal quantum of dynamic complexity, $h$ [7]). The vortex rotation initiated in the phase of reduction (squeeze) naturally continues (in the same 'direction') in the phase of disentanglement-extension, before restarting during the next reduction, etc., which gives the unceasing vortex rotation (spin), though maybe with a varying orientation of its axis. Note also that this physically real vortex motion of (many) spins, having a preferred average orientation of axes and taken rather in their 'extended' phase, provides the causal origin of the observed 'magnetic field' consistent with the 'laws of classical electrodynamics' [7]. Those effects of spin and magnetic field can be naturally included in our subsequent description of the global field-particle dynamics, but in this work we shall not consider them in order to concentrate on the more relevant properties of reduction and the related gravitational interaction.

We call the causally derived process of unceasing cycles of nonlinear (catastrophic) reduction-extension *quantum beat*. Here (and from now on elsewhere) 'quantum' is not the ordinary formal (and always 'mysterious') attribute, but the indication on the particular, *dynamical* and 'first-principle', origin of *discreteness* (or 'quantization') of *complex dynamics* of a system with interaction (the two protofields in our case). Being a priori spatially homogeneous and timeless, the system dynamics 'creates' the property of discreteness 'by itself', eventually because of the intrinsic *wholeness* of the *unreduced* interaction process (properly taken into account in the universally nonperturbative EP method), where 'everything is related to everything' and therefore arbitrary small, purely local motions cannot exist. This purely dynamical emergence of discreteness characterises, at the same time, the appearance of the elementary object, the elementary field-particle, represented by the same quantum beat and realising the first level of the autonomous structure formation process of the world 'constructed' from the two interacting protofields. It becomes clear [7] that the dynamic discreteness-quantization, together with the discovered causal uncertainty of choice among equally possible, but incompatible realisations, provides the causally complete, complex-dynamical extension of the 'quantum indeterminacy' and the related 'uncertainty relations' in the behaviour of elementary field-particles. The same result can be expressed in other words by saying that the quantum beat process provides the causal origin, and the detailed universal structure, of the 'wave-particle duality', thus completely demystifying and explicitly resolving this most intriguing problem of the standard 'quantum mechanics' (it is indeed 'irresolvable' within any unitary, single-valued description, as it was honestly recognised by Niels Bohr within his 'standard', or 'Copenhagen' interpretation of quantum behaviour). The 'wave' and the 'particle' indeed physically 'coexist' (or rather permanently *alternate*) within one and the same elementary object-process, being represented respectively by extended, 'intermediate' state-realisation of the system and many its localised, dynamically squeezed state-realisations. Moreover, the undular state is permanently *transformed* into the corpuscular one and back (as it should be), so that at a sufficiently high frequency of quantum beat[4] one will always observe a 'coexistence' of both dually opposed states/properties in practical measurements. From the point of view of the localised, 'corpuscular' state and aspect of the dualistic quantum beat dynamics, the latter can be conveniently presented as the unceasing *chaotic wandering* of the localised state which can be called *virtual soliton* for its irreducibly unstable character (such kind of 'thermal motion' *within* each field-particle extends the idea of 'hidden thermodynamics' of Louis de Broglie, see below, section 2.2). In summary, all the 'unsolvable' mysteries of (standard) quantum mechanics can now be causally and completely explained as being due to quite natural, universal (but certainly non-trivial) properties of the unreduced *dynamic complexity* of a (generic) interaction process. It is also clear why they *cannot* be consistently understood within *any* version of the canonical, single-valued approach, including the existing imitations of complexity [5,7].

It is clear from the described results that, as it should be expected, the elementary particles emerge from the protofield interaction together with the elementary spatial structure and temporal rhythm. The 'embedding' *space* and *time* of the world acquire a *physically real* and irreducibly *dynamical* basis (contrary to their purely abstract, 'mathematical' sense in the canonical science, including the standard 'relativity'): the elementary physical 'space point' is given by the dynamically squeezed state of the elementary field-particle, while the elementary 'time interval' is marked by emergence of consecutive reduction *events* within the corresponding quantum beat process (i. e. it is the *emerging* quantum beat *period*). We see that the space is naturally *discrete* (due to the *wholeness* of complex interaction dynamics), while time is naturally *irreversible* (due to the *causal randomness* of choice of each next spatial point, resulting from the dynamic redundance and thus eventually also from the same wholeness of the unreduced interaction process), and both properties causally *appear* as intrinsic features of the interaction process in an a priori strictly homogeneous and timeless system.[5] Below we provide these results with mathematical expression and show that the dynamical origin of causal space and time actually contains the intrinsically complete, physically transparent understanding of their (extended) *relativity* which originates thus from the *same dynamic complexity* as the 'quantum' behaviour of the object. On the contrary, any 'symmetry' between space and time, so persistently imposed by the canonical, unitary relativity, is definitely excluded from their causal extensions: space is a 'tangible', 'material' *structure* and thus indeed a physically real 'dimension', while time characterises the events of (unceasing) *emergence* of those structures and cannot *be* itself a real dimension. In particular, our causal time naturally and unceasingly *flows*, due to the permanent character of the quantum beat process and in accord with observations, whereas spatial changes as such need not be 'oriented' in a single direction (actually, they vary chaotically). As will be confirmed below, the property of 'relativistic' *mass-energy* of the elementary field-particle (and thus eventually any many-particle body) can be consistently defined as the *temporal rate, or frequency* (characterising the 'intensity') of the *spatially chaotic* quantum beat process.

---

[4]The frequency is indeed very high, of the order of $10^{20}$ Hz for the electron (see below).

[5] It is also important that quantum beat processes of *all* individual elementary particles of the World are *synchronised* in phase, up to its inversion [7], which ensures *universality* of the relative time.

There is no wonder that this intrinsic unification of dynamical quantization and causal extension of 'special' relativity of space and time actually includes also the causal, complex-dynamical extension of 'general' relativity, in the form of physically complete understanding of the *universal gravitation*, both in its Newtonian version and (extended) 'relativistic' modification. Before specifying below the mathematical expression for that unified picture, we outline now its qualitative aspect. The origin of the universal gravitational attraction between material objects implies the existence of many such objects, and in our 'emergent' description they start appearing as many individual quantum beat processes within the *same* system of two interacting protofields each of those individual processes being observed as an elementary field-particle with the dualistic 'quantum' properties described above. There can be eventually several different species of really elementary field-particles [7,11], each of them understood as a particular dynamical regime, or again realisation, of the protofield interaction (in particular, the *electron* can be considered as the simplest, or 'prototype', elementary field-particle resulting from the described protofield interaction process).[6] Each individual particle-process can be considered as a new object-realisation of the corresponding higher sublevel of complexity and those objects start interacting with each other, and that interaction develops according to the same universal FMDF scenario resulting in the complex-dynamical processes of realisation change that form a still higher level of complexity.

The world progressively emerging in this way from the originally homogeneous system of two interacting protofields can be figuratively described as the physically unified 'world sandwich' consisting from two protofield 'sheets' that form its two 'sides' and are dynamically 'pinned' against each other in many (changing) points by individual field-particles (instead of being homogeneously and statically 'stuck' to each other, as it would follow from the canonical, effectively one-dimensional analysis of the protofield attraction). It is clear then that the mentioned higher-level interaction between the individual field-particles can proceed by two different ways corresponding to its physical transmission through respectively e/m or gravitational side of the world sandwich. The mechanism of 'transmission' is generally straightforward in both cases and directly related to the quantum beat process of each field-particle: the 'mechanical' properties (like tension) of the protofield around each field-particle are considerably modified because of its contraction in course of each reduction phase, and therefore any other quantum beat process occurring at a finite distance from the particle will have modified characteristics (like frequency), which corresponds to a change of mass-energy referred to as particle 'interaction' (we naturally obtain thus the causal explanation of 'relativistic' equivalence between 'potential energy' of interaction and mass). Both physical ways of 'interaction transmission', or *action at a distance*, between the individual field-particles are always activated, due to *unceasing* character of their quantum beat processes within the *unique*, omnipresent protofield couple, and explain the causal origin of the observed e/m interactions and universal gravitational attraction, respectively. There is also some qualitative and quantitative difference between the two interaction types related eventually to difference in 'mechanical' properties of the two protofields and also to the fact that the world is asymmetrically 'displaced' towards the e/m side of the 'sandwich', so that only the e/m 'matter', in various its states, can be directly perceived within the world, whereas the gravitational protofield manifests itself only through the observable patterns of dynamic behaviour of the e/m 'material'.[7]

---

[6]This shows why and how the dynamic redundance concept permits one to avoid the artificial 'multiplication of entities' that *naturally* emerge instead in both real world development and its adequate description within the unreduced, intrinsically multivalued science of complexity. Contrary to this unique opportunity provided by the dynamic redundance paradigm, the canonical, single-valued science is obliged to take the entities it uses from either empirical observations (often ambiguous), or arbitrary 'theoretical' assumptions (despite the ban of 'Ockham's razor'), also only intuitively guessed in order to compensate, at least partially, the inefficiency of the inevitable trial-and-error technique of that unitary type of knowledge (we see that Ockham's conjecture can now be causally substantiated and extended within the dynamic redundance concept). The emergent character of the unified knowledge permits one also to avoid its artificial, simplified separation into 'theory', 'experiment' and their 'comparison', the latter being presented as a universal 'criterion of truth', despite its inevitably unrealistic, speculative character becoming more and more evident with growing complexity of the investigated phenomena.

Thus, the e/m mode of 'action at a distance' between individual field-particles proceeds by their *direct* relation through the common extended state of the e/m protofield ('intermediate system realisation'), in the 'extension phase' of the corresponding quantum beat processes. This connection between interacting quantum beat processes is *physically* realised by massless, but *dynamically* quantized undular perturbations of the extended e/m protofield known as *photons*, so that the revealed mechanism of the e/m action at a distance can be presented as the causally complete version of 'e/m interaction by *exchange of photons*' [7]. Due to this universal connection between individual quantum beat processes through the e/m protofields, their phases are *temporally* synchronised, so that there are *only two* possible types of temporal phase difference (up to $2\pi n$) between *any* two field-particles of the universe, 0 (in phase) and $\pi$ (in antiphase), which explains both the mentioned universality of the 'general' (relative) time of the World and the origin and discreteness of *electric charge* (it is a measure of the same quantum-beat complexity) with its *two, and only two* 'opposite' types (see [7] for more details). It is clear then that the e/m interaction between individual field-particles will depend on the likeness of their 'charge sign' (i. e. on the phase difference between the corresponding quantum beat processes).

---

[7]That is why the exact physical nature of the 'gravitational medium' inevitably remains much more ambiguous than that of the 'e/m medium', even though it can be gradually revealed within further studies of its manifestations in the detailed structure of the observed fundamental entities and interactions. In particular, what we call here 'gravitational protofield' can be closer to something like the omnipresent 'border of the (e/m) world' upon which it 'tries to fall', e. g. under the (attractive) influence of the 'other side' of the border, until the appearance of high enough 'forces of reaction'; or else the function of the gravitational protofield can be actually performed by an 'antipode (e/m) world', for example the antimatter world, dualistically related to (but separated from) our world [7]. It is important to emphasize, however, that the universality of our description, and the *physical reality* (extended causality) of the underlying entities, are generally preserved in *every* such situation which can eventually be clarified by summarising all the finer details of the expected complete picture of the observable world (or maybe by going, one day, to the outside of *this* world?!).

Contrary to such more direct way of action at a distance, its another mode, universal gravitation, transmitted through the gravitational medium (protofield) does *not* depend on phases of the participating quantum beat processes, but only on the 'average' increase of tension of the gravitational medium, which explains the existence of *only one* 'sign' of 'gravitational charge', the latter being reduced simply to the same mass-energy that is defined as the quantum-beat temporal rate and characterises the effects of ('relativistic') inertia (see also below). It is also clear why the universal gravitational interaction can take only the form of attraction: each field-particle tries to 'pull' the common gravitational 'blanket' to itself and therefore all the particles are pulled to each other, due to the finite elasticity (extendibility) of the gravitational medium. We see that, contrary to the 'active' (direct) photonic connection of individual quantum beat processes through e/m protofield, the gravitational medium plays the role of a 'passive' (or indirect) intermediator that transmits the tension between the field-particles, but does not directly relates them by transversal wave-like perturbations analogous to photons (the perturbations of the gravitational protofield can exist, but they are not directly 'attached' to their particle-sources 'situated' at the e/m side of the 'sandwich' and have rather longitudinal character). Nevertheless, the obtained causal origin of universal gravitation has a naturally *quantized* (dynamically discrete) character, so that our causal gravity is an *intrinsically quantum* phenomenon, even though in ordinary conditions (including the usual 'quantum' scales) we deal with quasi-classically large 'numbers of quanta' of gravitational attraction which therefore cannot be individually discerned. This irreducibly *quantum origin of gravity* is directly related to its interpretation as an *unceasing*, omnipresent, complex-dynamical *process* (of quantum beat), rather than a (usually) *stationary*, or uniformly changing, 'force' or 'curvature' (within the canonical approaches, including Einsteinian 'general relativity' and all its unitary modifications).

Before presenting a more rigorous expression of the outlined picture of unified dynamic complexity at the level of elementary world structures, it would be not out of place to mention the existence of a number of further important details and independent correlations contributing to its consistent completion and thus also, indirectly, confirming the proposed interpretation of universal gravitation. Thus, the other two interaction types between elementary particles, known as 'weak' and 'strong' interactions, can be consistently interpreted within the same unified picture as particular dynamic regimes, or 'realisations', within the same dynamical system of coupled protofields [7] responsible for short-range interaction processes and eventually originating from the corresponding direct connections between individual constituents (unobservable in pure form) of the e/m and gravitational protofields (within each protofield), respectively. This interpretation agrees with the observed manifestations of relation between the weak and e/m interactions. In general, we have two protofields, each of them serving as a source of the corresponding 'internal', short-range interaction and also as a way for transmitting a long-range 'action at a distance', which naturally gives $2 \times 2 = 4$ observed fundamental interactions. Note that contrary to the canonical approach dealing with artificial (mechanistic), and therefore basically unsuccessful, 'unification' of those 4 interactions, in the universal science of complexity (represented by the 'quantum field mechanics' at the considered lowest levels of complexity) we have the 'unified interaction' (together with the physically specified 'interacting system') from the very beginning, in the form of the coupled protofields, and then discover how all the particular 'interactions', but *also the interacting objects*, naturally *emerge* from that simple initial system *without* any additional 'postulates', or inserted physical 'entities'. This really 'first-principle' approach is the only way to the causally complete understanding of objective reality (thus proving the very *possibility* of such *intrinsic completeness* of a *developing* knowledge [7], which is a subject for serious doubts within the canonical science), and in particular it provides the clear 'guiding line' showing what and why can, or cannot, exist at the basement of being otherwise looking so 'mysterious', and therefore remaining open to so many arbitrary guesses, within the unitary science. Thus, the 'mysterious' existence of a 'redundant' number of building blocks of nature, in the form of several 'generations' of ever more exotic elementary particles, can be understood as an inevitable result of the same universal dynamic redundance of the unified protofield interaction which, being a *complex* dynamical process, *cannot* produce strictly the necessary minimum of objects (because then its complexity would be zero), which is quite similar to the origin of natural 'diversity of species' at various higher levels of complexity. The real, physical 'unification' of the 'constituent' interaction types permanently, *dynamically* occurs 'within' the elementary field-particles (and thus also during their mutual transformations, now also causally specified), i. e. at the scale of individual quantum beat cycles, as it is expressed by the (causally modified) Planckian units (see below). This shows already that one does not need to look for vague, indirect sings of a formal 'Grand Unification' in the extremely expensive, and as much ambiguous, experiments at 'super-high' energies *erroneously* imposed by purely mechanistic, senseless calculations of the unitary science (we develop the subject in section 3). It is sufficient, instead, to study in more detail the known particle and interaction species, which is quite a feasible task for the existing high-energy facilities, while the efficiency of their use can indeed be dramatically increased if one applies the causally completed picture of reality.

 Another series of independent correlations supporting that picture comes from a somewhat higher sublevel of complexity, that of the quantum behaviour of interacting field-particles of the same species. The quantum field mechanics provides a physically transparent interpretation of the canonical quantum 'postulates' about 'indistinguishability of individual particles' and 'Pauli's exclusion principle' surrounded with the usual aura of 'quantum mystery' in the canonical theory. Since an elementary particle is a complex-dynamical process of the *first* observable level of complexity involving just certain amount of the *same*, initially uniform protofield matter, it is clear that different participating parts of that matter cannot be observably distinguished from each other, the particles can (and actually permanently do) exchange them, they can be 'transformed' one into another, 'silently' exchange places with each other, etc. As to 'Pauli's exclusion principle', it simply expresses the existence of a fixed 'amount of (protofield) matter' within the reduced (corpuscular) state of each particle (for the fixed magnitude of the unique protofield coupling), so that no any other particle can take exactly the same state (position). This intuitively clear rule can be more rigorously specified as a manifestation of the universal complexity conservation law (since each particle-process contains the fixed total number of realisations). The *apparent* violation of this rule in the behaviour of bosons is actually due to their well ordered spatial arrangement within 'Bose condensates' impossible for fermions (further details can be found in ref. [7]). Note that the photons, contrary to the canonical interpretation, are quite different from independent, massive bosons. The photons, within our approach, are indeed much closer to 'linear' and therefore *always* extended e/m waves (perturbations of the e/m protofield); in particular, they do not contain any internal quantum beat process, possessing thus zero dynamic complexity and therefore zero mass (even though their 'passive' interaction with the gravitational protofield can play some role in their properties and introduce a 'weak' or 'hidden' complexity in their behaviour). As to 'quantized' character and 'particle-like' behaviour of photons in interaction processes, all such features can be reduced to the corresponding properties of the involved quantum-beat processes within the participating massive particles. While the measured 'electric field' results from (many superimposed) transversal oscillations of the e/m protofield, excited by quantum beat, the 'magnetic field' originates from many spin vortices, excited in course of reduction as described above, with partially ordered directions of their axes. This interpretation of the measured e/m field permits one to causally explain the experimentally known 'laws of electrodynamics' [7] and corresponds much closer to the original ideas of James Maxwell, than their subsequent mechanistic simplification within the scholar science (including Maxwell's 'ether' as the indispensable medium where the e/m waves are excited and which corresponds to our 'e/m protofield', whereas it was formally rejected in the standard 'relativity'). It becomes also clear that in the absence of dynamical reduction of the photonic wave giving rise, in particular, to the physical vorticity of spin, attribution of a non-zero spin to photons in the canonical theory has no physical sense and serves rather as formal simulation of the e/m interaction processes of the massive fermions [7].

## 2.2. Quantum beat dynamics of the globally moving elementary field-particle

Turning now to mathematical expression of the main observable manifestations of the quantum beat dynamics, we start with distribution of reduction probabilities for the *isolated* field-particle (exemplified by the electron). Each reduction event leads to formation of a particle-like, compressed state of the e/m protofield centred around an emerging 'position' (see eq. (6)) chosen by the system in a causally random fashion among many equally probable 'realisations' of a 'space point'. It follows that the corresponding probabilities, $\alpha_r = \alpha(x)$, of reduction events are *equal* for all possible 'points' of 'particle position' and therefore do not actually depend on the (emerging) space coordinate of realisations, $x$:

$$\alpha_r = \frac{1}{N_\Re} \quad (r = 1,...,N_\Re), \quad \sum_r \alpha_r = 1, \qquad (8a)$$

where $N_\Re$ is the total number of realisations for the field-particle, introduced above ($N_\Re \sim 10^2 - 10^3$ for the electron). Therefore the wave field of the isolated field-particle does not contain any measurable 'coordinate dependence', which is not surprising in view of the causally random, permanently changing 'positions' of the emerging 'space points'. When the latter become partially ordered (for example, in the case of moving, non-isolated field-particle, see below), the probabilities $\alpha_r$ are not equal any more and acquire the corresponding dependence on $r$ (and thus $x$):

$$\alpha_r(N_r) = \alpha(x) = \frac{N_r(x)}{N_\Re} \ (N_r = 1,...,N_\Re; \sum_r N_r = N_\Re), \quad \sum_r \alpha_r = 1, \qquad (8b)$$

where $N_r$ is the number of the 'elementary' system realisations within their $r$-th group, those groups being actually measured in experiment and playing the role of 'effective realisations'. Note that eqs. (8) can be considered as expression of the causally extended, generalised 'Born's probability rule' which follows, in its turn, from the dynamical boundary (initial) conditions for the general solution (see eq. (6)) [7,10].

By contrast to the (observed) spatial homogeneity of the isolated field-particle, the emerging temporal 'ticking' of its quantum-beat cycles should have some observable manifestations because it corresponds to a regular, 'one-dimensional' dependence. It is evident that two successive reduction events are separated by the elementary (smallest possible) interval $\Delta t$ of the emerging time. It should be proportional to the corresponding increment of a physical quantity, $\mathcal{A}$, adequately characterising the global change of the system between the two reductions and expressing the unreduced dynamic complexity of the field-particle (considered as a process): $\wp \Delta t = \Delta \mathcal{A}$, where $\wp$ is the coefficient. It is clear that $\wp$, being the discrete analogue of the time derivative of $\mathcal{A}$ ($\wp = \Delta \mathcal{A}/\Delta t = d\mathcal{A}/dt$), expresses the temporal rate, or 'intensity' of quantum beat and thus also characterises its complexity. For the isolated field-particle one can find one and only one physical property with such features, and it is the *inertial mass* of the particle. The quantum beat intensity fits well to the role of mass due to its self-sustained and spatially random character, since any attempt to change this state will evidently meet the effective resistance of the *already existing* (and self-preserving) *internal motion* (cf. the resistance to compression of a gas consisting from chaotically moving molecules). Thus $\wp$ is proportional to the rest mass, $m_0$, of the field-particle which should eventually be equivalent to its (rest) energy, $E_0$. At the same time the above expression for the temporal rate of the quantum beat process should be compatible with the relation between energy and action known from classical mechanics: $E = -\partial \mathcal{A}/\partial t$. Therefore we can put together all those correlations in a unique self-consistent structure by assuming that the above quantity $\mathcal{A}$ is indeed action (of Maupertuis) permanently changing (actually *diminishing*) in course of the quantum beat process. Action seems to be a suitable characteristic of the global system behaviour, and at the same time its proposed interpretation in terms of the highly nonlinear quantum beat process provides a new insight into the physical sense of this popular physical quantity, now becoming really universal (we continue specifying the involvement of action below). Taking into account the natural *quantization* (dynamical discreteness) of quantum beat, we can finally present the found basic relation between mass (energy), time, and action in the form:

$$E_0 = m_0 c^2 = -\frac{\Delta \mathcal{A}}{\Delta t} = \frac{h}{\tau_0} = h\nu_0 ,  \qquad (9)$$

where for the moment $c^2$ is simply a coefficient (to be properly specified later), $-\Delta \mathcal{A} = h$ is the quantum of action-complexity corresponding *physically* to one cycle of the quantum beat (nonlinear reduction-extension of the coupled protofields or one 'quantum jump' of the 'corpuscular state', or *virtual soliton*), $\Delta t = \tau_0$ is the *emerging* 'quantum of time' equal to one period, $\tau_0$, of the same cycle, $\nu_0 \equiv 1/\tau_0$ is the corresponding quantum beat frequency, forming the basis of the causal time concept [7,11] ($\nu_0 \sim 10^{20}$ Hz for the electron). The causally substantiated meaning of eq. (9) provides, in particular, the physically complete understanding of the resulting relation between $m_0$ and $\nu_0$, $m_0 c^2 = h\nu_0$, which was semi-intuitively postulated by Louis de Broglie and used for derivation of the famous expression for the 'de Broglie wavelength' [20] and further development towards the 'double solution' concept [21-24] (see also below). We see now that the mysterious 'oscillating (periodic) phenomenon' inexplicably 'attached' to the particle ('mobile') is the nontrivial reduction-extension pulsation of the quantum beat process containing the *essential* (and *uniquely* defined) *nonlinearity*, intrinsic *spatial randomness* and permanently *transforming* the localised 'particle' (or 'virtual soliton') into the extended wave field and back, this 'wave' thus being also naturally 'attached' to the 'particle'. The dynamically random and unceasing character of quantum beat is important also for the causal explanation of the *irreversible* and permanently *flowing* character of the emergent *time*, *t*, defined within the same eq. (9) as (up to an additive constant)

$$t = n\tau_0 = n \frac{h}{m_0 c^2} ,  \qquad (10)$$

with permanently growing integer *n*. This causally derived time is directly related thus to the unreduced dynamic complexity (nonlinearity and dynamical randomness) and therefore can be only superficially *simulated* by a (linear) 'operator of time' artificially endowed with the *formal* property of irreversibility within certain basically restricted interpretations of 'complexity' in the single-valued science.

When the field-particle isolation is violated by its interaction with other particles, the observed spatial homogeneity of its wave field (extended, or 'intermediate' realisation) in the isolated condition is replaced by a measurable 'coordinate dependence' causally emerging in accord with the nonuniform case of the 'probability rule', eq. (8b). This *global* (average) inhomogeneity of the particle wave field corresponds to its (global) *state of motion*, whereas the above homogeneous wave field of the isolated particle realises the (global) *state of rest*. The latter can be rigorously defined as the state of the field-particle (quantum beat process) with *minimum* dynamic complexity (as it is represented by mass-energy), while any *increase* of the quantum beat complexity over that (unique) minimum produces one of many possible states of motion [7,11]. In this way we obtain, within the quantum field mechanics, the rigorous and universal *definition of motion itself* not reduced to an ambiguous 'displacement' or 'position measurement' which always remains ill-defined itself. Complexity growth of the field-particle observed as its motion occurs physically by appearance of a partial *dynamical order* in the spatial distribution of reduction centres, even though each next reduction always happens in a causally random, probabilistic fashion (i. e. we *can* indeed predict the *probabilities* of different possible reduction events/realisations, and thus their *average*, or 'expected' *spatial distribution*, but *not the exact* occurrence of any future event). This average spatial order of the moving field-particle is described mathematically just by appearance of a 'coordinate dependence' of the 'attached' wave field of the 'main (intermediate) realisation' (indeed, since the quantum beat period $\tau_0$ is the shortest really existing period of time, corresponding to a very high frequency $v_0$, only the average tendency, or 'order', in the reduction centre distribution can actually be observed). This means that for the moving field-particle, contrary to the one at rest, the above introduced action-complexity depends now on both time and coordinate, and therefore one should distinguish between the total and partial time derivatives of action as they are involved in the above definition of mass (energy):

$$\frac{d\mathcal{A}}{dt} = \frac{\partial \mathcal{A}}{\partial t} + \frac{\partial \mathcal{A}}{\partial x}\frac{dx}{dt} = -E + pv ,$$

where $E$ is the total energy-complexity,

$$E = -\frac{\partial \mathcal{A}}{\partial t} ,$$

while the *momentum*, $p$, that characterises the emerging spatial order in the probability distribution of the moving field-particle and its *velocity*, $v$, are introduced in accord with the canonical relations:

$$p = \frac{\partial \mathcal{A}}{\partial x} , \quad v = \frac{\partial x}{\partial t} .$$

Taking into account the quantized character of the quantum beat dynamics, the relation of eq. (9) derived for the isolated particle should be replaced, for the general case of the moving field-particle, by [7,11]

$$E = mc^2 = -\frac{\Delta \mathcal{A}}{\Delta t} + \frac{\Delta \mathcal{A}}{\lambda}\frac{\Delta x}{\Delta t} = \frac{h}{T} + \frac{h}{\lambda}v = hN + pv , \qquad (11)$$

where

$$E = -\frac{\Delta \mathcal{A}}{\Delta t}\Big|_{x=\text{const}} = \frac{h}{\tau} = hv , \qquad (12)$$

$$p = \frac{\Delta \mathcal{A}}{\Delta x}\Big|_{t=\text{const}} = \frac{\Delta \mathcal{A}}{\lambda} = \frac{h}{\lambda} , \qquad (13)$$

$$v = \frac{\Delta x}{\Delta t} \equiv \frac{\Lambda}{T} , \qquad (14)$$

$\lambda \equiv (\Delta x)|_{t = \text{const}}$ is the *emerging* 'quantum of space', a minimum directly measurable (regular) space inhomogeneity characterising the elementary quantum field with complexity-energy $E$ ($> E_0$) and resulting from its global displacement (motion), $\Delta t = T$ is the 'total' period of nonlinear quantum beat of the field in the state of motion with complexity-energy $E$ ($N = 1/T$ is the corresponding quantum-beat frequency), $\Delta x = \Lambda$ is the 'total' quantum of space, while $\tau = (\Delta t)|_{x = \text{const}}$ is the quantum-beat oscillation period measured at a fixed space point.

It is not difficult to understand that the elementary space inhomogeneity $\lambda$ of the wave field of the moving field-particle, thus *dynamically* emerging in the underlying process of quantum beat, is none other than the 'de Broglie wavelength of the particle', $\lambda = \lambda_B = h/p$, which provides this rather ambiguous notion of the canonical quantum mechanics with a causally complete interpretation. The latter is closely related to the complex quantum-beat dynamics in the 'parent' system of protofields which necessarily includes the extended, undular state-realisation ensuring chaotic transitions between different corpuscular states of the system. The global motion of this complex-dynamical process in a definite direction corresponds to a partial order of the chaotic structure of the intermediate realisation physically reproducing the distribution of probabilities of corpuscular state-realisations. The 'uniform, rectilinear' motion of the field-particle corresponds to the observed, ordered part of the wave field in the form of a 'plane wave' with the de Broglie wavelength $\lambda_B$, even though this wave is not really as 'linear' as it appears in the formal description of observations and is closer, in reality, to a quasi-standing wave with regularly situated nodes between which the actually moving part of the field-particle, the corpuscular 'virtual soliton', performs its causally random jumps. In other words, the 'de Broglie wave' of a particle is a 'self-organised' structure of its transient extended state that can, however, maintain its coherence (repetitive regularity) due to the complex-dynamical 'synchronisation': other space-structure elements, with chaotically varying size and position, also exist within the particle wave field, but none of them, except for those with the size $(\Delta x)|_{t = \text{const}} = \lambda_B$ and coherent with the existing de Broglie wave, can actually survive as a repetitive, perceivable 'in average' structure. This physical picture corresponds to the partition of the total energy-complexity into the two parts described by eq. (11): the last part, $p\upsilon = h\upsilon/\lambda$, represents the regular, average component of the particle wave field properly characterising its global motion tendency, while the first part, $hN = h/T$, corresponds to the remaining 'purely' irregular deviations from that global tendency which therefore can be characterised by a temporal, rather than spatial, structure elements (even though a 'characteristic' size can still be attributed to this highly irregular wave-field component, in the form of the 'Compton wavelength', $\lambda_C = h/m_0c$, coinciding with the length of the elementary 'quantum jump' of the 'virtual soliton' [7,11]). The same partition of the total energy into averaged and irregular parts exists for any nonuniform type of the global motion, and their mathematical expressions can be further specified in relation to the particular nonuniformity in question. Note also that one should distinguish between the observed elementary space structure of the moving field-particle, $\lambda_B$, and the still more fundamental, actually the smallest ever emerging, spatial element from the lower sublevel of the same complex dynamics that characterises the size of the corpuscular state itself (it is given by the 'modified' Planck length defined in the next section and can also be considered as the 'smallest possible' Compton wavelength for a particle in this world).

The described partition of the complex dynamics of a moving field-particle cannot, by itself, completely specify the detailed character and observable manifestations of that dynamics. Further refinement should necessarily include a direct relation between the spatial 'density' (or typical 'size') of the emerging structure and temporal rate of its emergence, i. e. between the generalised momentum $p$ and energy $E$ defined respectively by eqs. (13) and (12). This complex-dynamical extension of the canonical *dispersion relation* can be obtained by elementary considerations [11] explicitly taking into account the existence of the experimentally 'hidden', but large part of the quantum beat dynamics, in the form of the described irregular 'wandering' of the virtual soliton around the average measured tendency. Since the latter corresponds to the global field-particle displacement with velocity $v$; whereas all the individual (hidden) virtual soliton motions within the e/m protofield are performed at the speed of light, $c$, it is clear that the proportion, $\beta$, (relative number) of the virtual soliton jumps falling within the global motion tendency is equal to $v/c$. Therefore, the elementary *global* advance of the quantum beat process, $\lambda$, is actually performed not during the elementary jump period, $\tau$ (which would give $\lambda = c\tau$), but during the time period of $\tau' = \tau/\beta = \tau c/v$ (in average) taking into account a usually large number ($1/\beta = c/v$) of *local* (irregular) deviations of the virtual soliton around the explicitly observed regular tendency. Since any jump is always performed with the speed of light, we obtain $\lambda = c\tau' = \tau c^2/v$. Inserting here the definitions of $\tau$ and $\lambda$, eqs. (12) and (13), we obtain the desired dispersion relation in the form

$$p = E \frac{v}{c^2} = mv , \qquad (15)$$

where $m \equiv E/c^2$, now by rigorously *derived* definition, which justifies, in particular, the exact relation between the (causally specified) mass and energy. We see now that this familiar 'relativistic' relation, seeming to be especially simple in the universal form, $p = mv$, actually hides in it all the unreduced complexity of the quantum beat dynamics (or, in general, realisation change process at arbitrary level of complexity). There is no wonder that not only it provides the causally complete extension of the corresponding relation from the canonical 'relativity', but can also be considered (together with the above causal notion of mass) as the rigorous derivation of Newton's laws of motion (the first two of them are obtained by taking time derivative of eq. (15)) which therefore also eventually result from the internally *complex* dynamics, despite their non-complex *axiomatic* formulation and *external* manifestation. The dispersion relation thus derived can be used now for reproducing the canonical expression for the de Broglie wavelength, but now derived really from the first principles:

$$\lambda \equiv \lambda_B = \frac{h}{mv} . \qquad (16)$$

Being actually the explicit expression of the internally continuous, but nonuniform, character of the quantum beat dynamics realising the unceasing transformation of the extended 'wave' into the localised 'particle' and back, the dispersion relation can be presented also as the causally complete extension of the 'phase accord theorem' used by de Broglie [20] in his semi-intuitive justification of the expression of eq. (16) and further development of the double solution concept (the details can be found in refs. [7,11]).

Using the dispersion relation of eq. (15) in the complex-dynamical partition of the total energy, eq. (11), we obtain the relation between the different time measures, $\tau$ and $T$, of the *same* quantum beat period:

$$\tau = T\left(1 - \frac{v^2}{c^2}\right). \tag{17}$$

As is seen from eq. (11), the time measured in units of $T$ and corresponding to the *total* temporal derivative of action-complexity is the 'internal', or 'proper' time of the *moving* field-particle (and eventually of a body composed from many elementary particles), which agrees with the fact that it represents the purely irregular part of the total quantum-beat energy (while the regular global tendency cannot contribute to this internal time determining the temporal behaviour of any internal process within the moving body). The other temporal measure of the same complex-dynamical process measured in units of $\tau$ and corresponding to the *partial* temporal derivative of action-complexity (taken at a fixed coordinate value, i. e. in the 'rest frame') is the 'external' time of the moving field-particle (or body) expressing, in particular, the temporal rate of the regular global oscillation of the quantum beat, the (causal) de Broglie wave observed from the state of rest. The physical, causal origin of the two temporal periods, $\tau$ and $T$, and the difference between them described by eq. (17) is clear: in order to ensure the global motion of the field-particle its quantum beat should increase the frequency, $v = 1/\tau$, of its oscillation with respect to its value in the state of rest, $v_0 = 1/\tau_0$, at the expense of the regular tendency growth, which inevitably leads to diminishing of the frequency, $N = 1/T$, of quantum jumps within the purely irregular part of dynamics explaining the appearance of the period $T > \tau_0 > \tau$. The latter rule of energy 'redistribution' between the average and irregular tendencies is a manifestation of the universal 'complexity conservation law' [7] which in this case is simply reduced to the fact that the quantum beat dynamics of the uniformly moving field-particle is always driven by the same, unchanged coupling (attraction) between the two protofields, irrespective of the current velocity $v$ of the global motion, whereas greater velocities correspond to greater *proportion* (measured by $\beta = v/c$) of the global, 'ordered' tendency with respect to 'purely irregular' part of the process (even though *every* single jump within *each* of the two tendencies proceeds in a *causally random* fashion and can have as *many* as $N_\Re$ incompatible results, each of them predictable only by its probability). The conservation of the basic interaction complexity can be expressed also by an additional, easily verifiable [7,11] relation between the three frequencies or periods:

$$Nv = (v_0)^2, \tag{18a}$$

or
$$T\tau = (\tau_0)^2 . \tag{18b}$$

We can use these relations in order to transform the expression of the described causal time relativity by eq. (17) into an equivalent relation between the directly comparable flows of the 'proper' time in the states of rest and motion, $\nu_0$ and $N$ (or $\tau_0$ and $T$):

$$N = \nu_0 \sqrt{1 - \frac{v^2}{c^2}} \quad \text{or} \quad T = \frac{\tau_0}{\sqrt{1 - \frac{v^2}{c^2}}} , \tag{19}$$

which reproduces the canonical effect of 'relativistic time retardation in a moving frame', but now provided with a physically complete, *really* causal explanation *irreducibly* based on the *complex dynamics* of the quantum beat. Another combination of eqs. (17) and (18) provides the relation between the partial (oscillatory) temporal rates of quantum beat processes in the states of rest and motion,

$$\tau = \tau_0 \sqrt{1 - \frac{v^2}{c^2}} \quad \text{or} \quad \nu = \frac{\nu_0}{\sqrt{1 - \frac{v^2}{c^2}}} , \tag{20}$$

which is equivalent, in view of eqs. (9), (11), (15), to the (now causally extended) 'relativistic increase of mass' of the moving field-particle (or a many-particle body):

$$m = \frac{E}{c^2} = \frac{m_0}{\sqrt{1 - \frac{v^2}{c^2}}} . \tag{21}$$

Finally, the causally extended version of the 'relativistic contraction (of length)' of the moving particle/body can be demonstrated simply by multiplication of the first relation of eq. (20) by $v$:

$$l = v\tau = v\tau_0 \sqrt{1 - \frac{v^2}{c^2}} = l_0 \sqrt{1 - \frac{v^2}{c^2}} . \tag{22}$$

Thus, we have causally *derived*, within the same uninterrupted first-principle consideration, the main results of the canonical relativity, but now liberated from any abstract, ambiguous 'philosophy' and pseudo-scientific mystification and provided instead with much additional knowledge about the underlying complex-dynamical processes which naturally, inseparably *unify* them with the equally causal explanation of *quantum* behaviour of the *same* system. The major effect of 'special relativity', the dependence of spatial scales and time flow on the system motion, remaining basically unexplained within the canonical theory, is now causally specified as being due to explicit *emergence* of the *physically real* space and time in the complex, *essentially* nonlinear and *redundant*, dynamics underlying the *externally* observed tendency of regular (global) motion. This result only confirms the evident *necessary* condition of the genuine causality in any part of knowledge: in order to explain the behaviour of an entity (space and time in this case), one should have the *causally complete* interpretation of the *origin* of that entity itself. It is sufficient to compare the obtained physically complete *understanding* of time and space relativity with the restriction of its canonical version in order to see that Einsteinian theory is *objectively* nothing more than a *mechanistic*, artificially arranged *interpretation* of the classical relations of Lorentz (developed also by Poincaré in the direction of relativity) which unfortunately succeeded to be imposed as an *absolute* truth and a logically 'perfect theory'. It is based on 'probable', but basically inexplicable *postulates*, and *not* even about nature, but about *abstract laws* of nature; the postulates of a characteristic pseudo-philosophical flavour (like 'principle of relativity') were simply *adjusted* to the *known* mathematical result (Lorentz transformations accompanied by preliminary interpretation of Poincaré), in a tricky but basically evident fashion. The canonical relativity was thus imposed to the scientific community in a purely subjective (though indeed trickily arranged) way that denied even any possibility of further explanation (contrary to qualitatively very similar, and now *internally* related, departures from causality in quantum mechanics which were quite readily discussed by the *same* Einstein and the whole scientific community ever since then). This disgraceful interruption of a much more profound (and thus necessarily less 'quick') way of development, initiated by the natural adherents of causal thinking at the beginning of the century, has led the whole development of the fundamental physics in the wrong way of strictly unitary, mechanistic and formal, thinking ending with the current impasse (this happened despite the well-known, causally substantiated objections of many prominent scientists just after the appearance of Einsteinian 'theory'). The modern stagnating and fundamental ruptures between quantum and classical physics, quantum mechanics and relativity, field theory and particle physics, as well as many other untreatable 'difficulties' of the canonical science, all result from that 'seducing' deviation from the appearing causality towards superficial trickery of formal symbols which was intentionally imposed by the ones and accepted too easily by the others at the beginning of the 'century of liberation', *the last* century (see also sections 1 and 4).

It is not surprising that the mechanistic 'mixture' between space and time in a 'four-dimensional manifold of space-time', constituting another 'great revelation' and the whole underlying 'paradigm' of the canonical relativity, definitely shows now its *completely mistaken*, and misleading, essence. The above physically complete interpretation of the quantum field mechanics clearly proves that *only* space constitutes physically real 'dimensions' (or 'degrees of freedom' of the emerging *structure*), while time *cannot* be a tangible dimension (structure), since it marks the unceasing series of *facts* (*events*, or actions) of emergence of spatial structure elements in course of essentially nonlinear, and therefore nonuniform, process of 'realisation change' (or 'quantum beat', in this case). This rigorously derived result explains, contrary to the imitations of the standard relativity, why time naturally and irreversibly *flows* in one 'direction', and why our perceived space has three dimensions (this is because it is explicitly 'woven' as a dynamic entanglement of the three initial entities, two protofields and their 'interaction' [7,11]). It is also quite natural that the complete *physical understanding* of the entity of *elementary particle* as the complex-dynamical *process* of quantum beat in the system of interacting protofields is obtained *simultaneously* with the causally complete interpretation of space and time; it *should* be so *par excellence* in *any* really consistent description of the fundamental levels of the world.

Another main, this time indeed authentic, result of the canonical special relativity is the famous equivalence between mass and (total) energy, the two quantities being related by a coefficient equal to $c^2$. Note, however, that neither each of these entities (i. e. mass and energy), nor their proportionality, nor the value of the coefficient do not obtain any physical explanation in the standard theory. Instead of this, it proposes a series of purely mathematical and subjectively justified 'guesses' about the 'role of Lagrangian' and its 'probable' form, ending up with the corresponding axioms that do not even involve, this time, any 'philosophical' speculation [25,26]. The famous relation, $E = mc^2$, announced as the 'greatest' and of course irreproachable 'achievement' of the 'most perfect' theory, is then formally deduced starting from those unexplained assumptions with the help of the equally unexplained tools borrowed from classical mechanics. The causally complete, rigorous derivation of the same relation within the quantum field mechanics described above demonstrates once more the difference between the mechanistic manipulation of the unitary science around the formal, quantitative 'coincidence with experiment' and the intrinsically complete understanding of reality in the science of complexity. In particular, we *obtain* (see eqs. (9), (11)) the physically consistent meaning of the concept of mass-energy as the temporal rate of the spatially chaotic process of realisation change (reduction-extension of the interacting protofields, or quantum beat) *inevitably* unified with the simultaneously emerging causal interpretations of time, space, and elementary particle (for mass is a 'very' intrinsic property of matter-forming particles). The exact law of proportionality between *E* and *m*, including the exact value of the coefficient involved ($c^2$), is rigorously derived within the extended version of the 'relativistic dispersion relation', eq. (15), which inevitably reveals the nontrivial underlying physics of chaotic virtual soliton wandering naturally decomposing into the explicitly observable 'average' (global) tendency and hidden irregular deviations from it (at this point the intrinsic connection between relativity and quantum behaviour becomes especially evident). There is no wonder that the same, dynamically emerging relation provides the unique rigorous substantiation and ultimately profound meaning of Newton's laws of motion (actually also postulated in the canonical classical mechanics) and their relativistic extension. We note in this connection that, in view of generality of the problem formulation and analysis, the causal relativity outlined here is directly extendible to any higher level of complexity [7], which is important for the consistent picture of the whole world dynamics.

The direct, inseparable relation between causally extended relativity and quantum mechanics at the (lowest) levels of elementary particles can be summarised in the refined expression for the fundamental mass-energy partition into averaged and irregular tendencies, eq. (11), which now takes into account the dispersion relation, eq. (15), and the causal time retardation effect, eqs. (18), (19):

$$E = h\nu_0 \sqrt{1 - \frac{v^2}{c^2}} + \frac{h}{\lambda_B} v = h\nu_0 \sqrt{1 - \frac{v^2}{c^2}} + h\nu_B = h\nu_0 \sqrt{1 - \frac{v^2}{c^2}} + \frac{m_0 v^2}{\sqrt{1 - \frac{v^2}{c^2}}},$$

(23)

where $h\nu_0 = m_0 c^2$ (eq. (9)) and we introduce *de Broglie frequency*, $\nu_B$, defined as

$$\nu_B = \frac{v}{\lambda_B} = \frac{pv}{h} = \frac{\nu_{B0}}{\sqrt{1 - \frac{v^2}{c^2}}} = \frac{v^2}{c^2}\nu, \quad \nu_{B0} = \frac{m_0 v^2}{h} = \nu_0 \frac{v^2}{c^2} = \frac{v}{\lambda_{B0}}, \quad \lambda_{B0} = \frac{h}{m_0 v}. \quad (24)$$

According to the first relation from eqs. (24) and the corresponding physical picture of the globally moving field-particle outlined above, the de Broglie wave it describes is a *physically real* undulation, compatible with the ordinary connection between its wavelength and frequency, despite — or rather *due to* — a nontrivial complex-dynamical process 'summarised' by this 'average' wave that can alone directly appear in observations (it is described by the second summand in eqs. (23)). The dualistic, purely irregular tendency in the quantum beat dynamics of the moving field-particle is described by the first summand in eqs. (23) and can be considered as the causally complete extension of another concept advanced by Louis de Broglie, the 'thermodynamics of the isolated particle' [27-29] (naturally generalised to moving particles and their ensembles, within our approach). Indeed, the dynamical redundance paradigm reveals the irreducible source, and meaning, of causal, purely dynamic randomness in *any* really existing dynamical system and therefore not only provides the rigorous and *universal* substantiation for any cases of 'thermodynamics' and 'statistical physics', but also shows that *every motion* is inevitably *chaotic* (or 'stochastic') in its *basic* elements (realisation change) and thus *unpredictable* in at least some details. In the case of quantum beat dynamics, the 'thermodynamical' character of *all* virtual soliton jumps leads to the causal substantiation of the concept of 'total', or 'relativistic' mass (energy) of the particle which naturally incorporates — and now we really understand why — any part of dynamic complexity of the system, including its potential form known as 'potential energy' [7,11]. This causally complete interpretation of the *total* inertial mass of the particle extends and simultaneously simplifies the idea of the 'variable mass of a particle' acquired in its permanent interaction with a hypothetical source of chaoticity, 'subquantum medium' that plays the role of a 'hidden thermostat' in the original de Broglie's concept. In our approach we *obtain* from the first principles the source of chaoticity as the *naturally emerging* dynamical redundance of 'ordinary' interaction process in the simplest system of well specified physical entities (fields), so that the resulting *phenomenon* of dynamical chaos itself replaces the additional, now unnecessary *entity (object)* of 'hidden thermostat'.

The obtained 'relativistic energy partition', eqs. (23), is the differential relation between temporal and spatial changes of the dynamical 'field' of action-complexity $\mathcal{A}(x,t)$. As usually, the same relation can be presented in the integral form describing the progressive change of action in course of the discrete 'quantum jumps' of the virtual soliton (or 'realisation change', in the general case). Without going into detailed description of the obtained interpretation, we note only that it extends and generalises the 'least action principle' of classical mechanics and the related 'Lagrangian' approach to presentation of dynamics [7,11]. Within this extension, also predicted by Louis de Broglie, the 'virtual trajectories' of the system entering the formulation of the *postulated* principle of least action are replaced by the *real* chaotic wandering of the system between its realisations (in particular, between different space points in the case of quantum beat within the field-particle) as it is *rigorously deduced* within the above picture. The permanent realisation change is actually governed by the conservation of complexity of the system, which means that the mechanistically limited, abstract 'principle of *minimisation*' (or, in general, 'variational principle') of the canonical science is replaced by the *universal symmetry* (principle of conservation) of complexity causally substantiated above and directly extendible, together with the whole description, to *any* type of objects (level of complexity). The qualitative appearance of the observed dynamics depends on the relative magnitude of 'quantum jumps' with respect to the characteristic accessible space: relatively small (short-distance) jumps correspond to 'localised', 'trajectorial', 'classical' type of behaviour, whereas relatively large (long-distance) jumps can produce only a 'smeared', very indistinct 'trajectory' and correspond to a 'nonlocal', or 'wave-like' ('undular') behaviour.[8] The actual choice between the two regimes in each particular case depends on the nature of the interacting parts of the system and the interaction parameters. Thus, the typically 'quantum' behaviour of the elementary particles is due to the homogeneous character of the interacting protofields which does not impose any special limitation on the magnitude of each jump, that is the realisation probabilities are equal, or comparable (for the moving particles). However, if we consider the simplest *bound* system of two elementary particles attracted to each other, then it becomes clear that there is a big chance of emergence of a quasi-localised, *classical* type of behaviour in such system because the probability for the *independent* particle-processes to perform *random* jumps to a *larger* distance *remaining bound*, and thus in a highly *correlated* manner, is relatively small [7,11]. This natural emergence, and interpretation, of the *truly* classical, trajectorial behaviour already in the isolated elementary, and thus microscopic, bound systems (as opposed to the 'semi-classical', but still completely delocalised dynamics studied in the canonical theory, cf. [16]), without any ambiguous 'decoherence' due to an arbitrary 'external' influence, is an important additional correlation of the quantum field mechanics with the observed reality, confirming its consistency.

In accord with this picture of complexity development to its higher levels, the function of *Lagrangian* obtains its causal interpretation and universal extension in our approach, since it is none other than the 'purely irregular', 'thermal' part of the total energy (the first summand in eqs. (11), (23)) taken with the negative sign and mathematically expressed, in accord with the canonical prototype, as the *total* time derivative of our causal action-complexity, whereas the total energy is given by *partial* time derivative of action, taken with the negative sign (see eq. (11)). Comparing now eqs. (11) and (23), we *obtain* [7,11] the expression for the 'relativistic Lagrangian' which is only mechanistically 'fitted in' (and thus, postulated) by the standard relativity [25,26]:

$$L = \frac{\Delta \mathcal{A}}{\Delta t} = pv - E = -hN = -m_0 c^2 \sqrt{1 - \frac{v^2}{c^2}} \;. \tag{25}$$

---

[8]Thus, the so-called 'scars' observed in chaotic quantum systems and explained by some other, rather ambiguous, reasons within the canonical, single-valued concept of quantum chaos can well be an example of such extremely thick 'chaotic trajectory' of a quantum system at the level of interacting field-particles (see refs. [7,8] for more details and references on the single-valued and multivalued concepts of quantum chaos).

It is also clear from the above consideration that this purely irregular part of dynamics described by the Lagrangian is responsible for the internal, 'relativistically retarded' time $(T,N)$ of the moving system naturally emerging as the sequence of spatially random events/realisations of the irregular tendency (see eqs. (11), (19)). Together with other results, this interpretation of the Lagrangian can be directly generalised to any level of (complex) world dynamics. In this way we obtain the causally substantiated and considerably extended (actually universal [7]) version of Lagrangian approach of classical mechanics, which completes the above substantiation of Newton's equations and specifies the causal, physical origin of the related notions of 'trajectory' and 'material point'. Note that the classical system with well-localised trajectory that emerges starting from the elementary bound system as described above, is then again subjected to causal indeterminacy of the same origin (dynamic redundance), which leads to 'blurring' of expected trajectory, when a nontrivial interaction with other classical bodies appears at this higher level of complexity. It is very important for the consistency and universality of the whole picture that the basic points of analysis and its results are *reproduced* at *each* higher level of complexity within the *same* FMDF description demonstrating thus the natural, internally *continuous* (though very *inhomogeneous*) development of our *presentation* of reality in *exact correspondence* with the *actual, physical* emergence and development of elements of reality, thus remaining always completely specified. This is a quite new, dynamically realistic sense of 'extensions' and 'generalisations' in the unreduced science of complexity, as compared to purely mechanistic 'coincidence' in the unitary science, sometimes provided with a number of basically incomplete 'guesses'.

    The fundamental partition of the total energy of the field-particle, eqs. (11), (23), (25), acquires thus the intrinsically complete sense of the related causal interpretation directly extendible to arbitrary levels of complexity. In particular, this energy-complexity partition can be expressed, within a rigorously specified analogy, in terms of (first-principle) thermodynamics, if we just designate, for example, the total energy, $E$, as enthalpy, $\mathcal{H}$, the Lagrangian with the reversed sign as the 'internal energy', $-L = \mathcal{U}$, and the 'global translation energy', $pv$, as the '$\mathcal{PV}$ term' (where now $\mathcal{P}$ is pressure and $\mathcal{V}$ is volume): $\mathcal{H} = \mathcal{U} + \mathcal{PV}$. Still more interesting thermodynamical analogy can be obtained for the corresponding relation between the changes of quantities from eqs. (11), (23), (25) which is equivalent to the rigorously obtained and causally understood 'first principle of thermodynamics': $\Delta \mathcal{W} = Q + A´$, where now the total (internal) energy change $\Delta \mathcal{W} = \Delta E$, the 'internal heat' (or 'thermal energy')

change $Q = -\Delta L$, and the 'work performed over the system' $A´ = \Delta(pv)$.[9] It is not difficult to see that this causal 'first law of thermodynamics' does not differ from the relativistic generalisation of the 'second Newton's law' that has been causally substantiated above within the same relation, as well as any dynamics does not differ now from thermodynamics: due to the inevitably emerging dynamic redundance, *any* nontrivial system with interaction behaves in a *causally random* fashion which can be described as a (possibly nonequilibrium) thermodynamical behaviour, *sometimes* having the *external* appearance of the canonical 'smooth', 'Newtonian' dynamics. Due to the universality of description presented above this thermodynamical interpretation can be directly, and rigorously, extended to higher levels of complexity including properly 'thermodynamical' systems, which provides the causally complete substantiation for thermodynamics itself involving intrinsic temporal irreversibility helplessly sought for in the canonical science, etc. One can see also that the main 'fundamental principles' postulated in different domains of the canonical science, such as 'least action principle' in mechanics, 'Fermat's principle' in optics, 'principle of relativity' and the laws of thermodynamics are now rigorously substantiated, causally extended, and naturally merge within the single complexity conservation law (or universal symmetry of complexity) taking many different forms without the necessity for any additional substantiation in each particular case (see ref. [7] for more details). Note that the quantum mechanical partition of the total energy into the 'thermal' (irregular) and 'translational' (global) motion parts, the thermodynamical analogy, and the idea of 'unification of fundamental principles of physics' were proposed by Louis de Broglie within his unreduced 'nonlinear wave mechanics' also known as the theory of double solution (and hidden thermodynamics) [20-24,27-31]. The quantum field mechanics, and the universal concept of dynamic complexity in general, provide the essential causal completion of this approach through addition of the important 'missing link', qualitatively new phenomenon (and the adequate concept) of dynamic redundance that takes the form of irreducibly complex quantum beat process at the lowest levels of the world dynamics.

It remains for us to specify the quantitative presentation of the undular, nonlocal aspect of the field-particle dynamics in its inseparable unity with the corpuscular aspect already qualitatively substantiated above. It becomes clear, within the above picture of the quantum beat process, that the causal, physically real extension of the canonical, purely abstract 'wavefunction' is provided by the extended, 'intermediate' (or 'main') realisation of the quantum beat, in the form of the randomly changing, but partially ordered, wave field excited within the e/m protofield by its spatially chaotic reductions to the gravitational protofield. The omnipresent, and purely causal (dynamic), randomness of this extended wave field, the unceasing events of its reduction to the corpuscular state, and the irreducible role of the (hidden) gravitational protofield explain the 'strange' properties of the wavefunction that could not permit any its realistic interpretation within the unitary (single-valued) versions of the canonical quantum mechanics. In the quantum field mechanics we can, instead, *consistently* describe the wavefunction as the 'amplitude' of a *physically real*, tangible wave field (actually, the e/m protofield *perturbed* by its coupling to the gravitational medium) due to the *autonomously* maintained, dynamic *randomness* of its structure which is also permanently *transformed* into the corpuscular state and back demonstrating thus the irreducible, and also physically real, wave-particle duality. The actually observed, average structure of the wavefunction of a moving field-particle is reduced to the causally extended, 'dynamically standing' de Broglie wave described above, while it also contains the chaotically changing, *purely* irregular component with a characteristic length given by the Compton wavelength, $\lambda_C$. The most disputable 'Born's rule of probabilities', simply postulated in the ordinary description and providing just the key link between the abstract formalism and the real measurement results, is causally derived in the multivalued description from the 'dynamical boundary (initial) conditions' having a transparent physical interpretation [7,10]: the (causally defined) probability of reduction (dynamical squeeze) of the extended wavefunction, i. e. perturbed e/m protofield, towards a location, driven by its attraction to the gravitational protofield and creating the observed corpuscular state-realisation at that 'point', is proportional to the 'density' (or 'intensity') of the wavefunction around the location (cf. eqs. (8)). We shall denote the amplitude of the causally extended wavefunction of the quantum field mechanics by the same symbol, $\Psi$, as the 'state-function' of the initial system of interacting protofields, eq. (1), since these two entities represent the same physical object, but considered at the neighbouring sublevels of complexity: the wavefunction describes the specific structure of the state-function appearing in the process of the protofield interaction together with the elementary field-particle(s) and also elementary structures of this new sublevel emerging from interaction between the field-particles. Correspondingly, the observed wavefunction structure, representing also the distribution of corpuscular-realisation probabilities, is described by its dependence not on the arbitrary combination of (actually unobservable) 'coordinates' of the interacting protofields ($q,\xi$) as it is the case for the basic world state-function, eq.(1), but on the *emerging* spatial coordinate, $x$, of the dynamically formed corpuscular state which is a special, physically real *entanglement* of the fields of $q$ and $\xi$ (see eq. (6), where $r = r(x)$): $\Psi \equiv \Psi(x)$, and $\alpha(x) = |\Psi(x)|^2$, where $\alpha(x)$ is the probability density to find the particle at the point $x$. In this way we arrive also at the natural emergence of the *physically real* system configurations (= realisations = 'space points', in the simplest case) and their *dynamically complete* set provided with causal, intrinsic probabilities [11] which replace the *formally* (or empirically) introduced 'configurations' and their *abstract* 'space' in the unitary approaches constituting one of the known aspects of incompleteness of the canonical quantum mechanics.

---

[9]Within the same analogy, one can also introduce the 'internal temperature', $T$, of a moving field-particle according to the evident relation $kT = -L = m_0 c^2 \sqrt{1 - \frac{v^2}{c^2}}$, where $k$ is the Boltzman constant.

    Although the causally extended 'rule of probabilities' expresses the dynamical relation between the intermediate, delocalised realisation of the wavefunction and the 'regular', localised realisations (corpuscular states of the quantum beat process), this intrinsically continuous connection deserves a more detailed representation taking the form of the causal version of '(Dirac) quantization rules' which are simply postulated

in the canonical interpretations of quantum behaviour. Since the causal 'field of action', $\mathcal{A}(x,t)$, describes the temporal evolution of the realisation probability distribution, $\alpha(x)$, of the corpuscular state-realisations, and the wavefunction, $\Psi(x,t)$, characterises the amplitude distribution of the intermediate realisation permanently transformed *dynamically* (i. e. in a *locally continuous* fashion) one into another, it is clear that the formal relation between the two distributions should be generally close to proportionality, with the coefficient representing the characteristic value of action, that is Planck's constant, $h$: $\mathcal{A} = h\Psi$. This general connection can be further refined, if we take into account the detailed sequence of stages within the quantum beat cycle transforming the (chaotic) 'wave' into the 'particle' and back. Such elementary cycle of the quantum beat process can be presented as a combination of the 'wave-particle transformation', in the form of the protofield squeeze-extension, and the 'quantum jump' of the forming 'virtual soliton' to a new position. The first stage can be characterised by the corresponding change of the wavefunction, $\Delta\Psi$, whereas the second one is described by the increment of action, $\Delta\mathcal{A}$ (in reality both these 'stages' and the respective increments are entangled in an inseparable, holistic mixture of quantum beat, but these details cannot be resolved in observations). Since after each cycle the system returns to the same general state (or, in other words, its total dynamic complexity remains the same, according to the complexity conservation principle), both type of change, expressed in terms of action-complexity, should exactly compensate each other, that is $\Delta\mathcal{A} = -h\Delta\Psi/\Psi$, where $h$ and $\Psi$ are the characteristic values of action and wavefunction respectively 'remaining constant' during the change of the dual quantity [7,11]. In order to take into account the particular complex-number presentation of the wavefunction, one should add a numerical coefficient, $i/2\pi$, into this relation, after which it takes the form

$$\Delta\mathcal{A} = -i\hbar\frac{\Delta\Psi}{\Psi} . \tag{26}$$

Inserting this expression into causal definitions of energy and momentum, eqs. (12), (13), determined by $\Delta\mathcal{A}$, we obtain the 'quantization rules' canonical in their form, but provided now with the well-specified realistic basis [7,11]:

$$p = \frac{\Delta\mathcal{A}}{\Delta x} = -\frac{1}{\Psi}i\hbar\frac{\partial\Psi}{\partial x} \quad , \quad p^2 = -\frac{1}{\Psi}\hbar^2\frac{\partial^2\Psi}{\partial x^2} , \tag{27}$$

$$E = -\frac{\Delta\mathcal{A}}{\Delta t} = \frac{1}{\Psi}i\hbar\frac{\partial\Psi}{\partial t} \quad , \quad E^2 = -\frac{1}{\Psi}\hbar^2\frac{\partial^2\Psi}{\partial t^2} , \tag{28}$$

where the wave presentation of higher powers of $p$ and $E$ properly reproduces the wave nature of $\Psi$ [7]. The obtained result shows that the abstract 'operators' axiomatically attached by the canonical quantum theory to the 'observable quantities', like $p$ and $E$, realise but a highly simplified, effectively one-dimensional simulation of the underlying complex-dynamical processes. The same is true for the 'operators of creation and annihilation' which mechanistically simulate the nonlinear transformation between the 'corpuscular states/realisations' and the extended wavefunction ('intermediate realisation') (in particular, their axiomatically fixed 'commutation relations' simulate the complex dynamics of one cycle of such transformation in the process of quantum beat).

The canonical 'wave equations' postulated in the standard quantum theory can now be rigorously *obtained* by substitution of the causally substantiated quantization rules, eqs. (27), (28), into corresponding expressions for rates of change of action-complexity. Thus, the simplest forms of both Klein-Gordon and Dirac equations can be obtained by using the 'relativistic' expressions for the total energy-complexity, eqs. (11), (23), while the Schrödinger equation is obtained either as a non-relativistic limit of those equations, or by the direct insertion of the quantization rules into the non-relativistic dispersion relation, $E = (p^2/2m_0) + V(x,t)$ (see refs. [7,11] for more details). Note that the causal versions of the wave equations thus derived reveal the detailed, complex-dynamical basis of the real physical processes they describe, thus completely eliminating the aura of mystery around the canonical, postulated versions. In particular, the fundamentally discrete, 'quantized' character of the measured quantities described by the wave equations and actually postulated in the standard interpretation is transparently explained by the discrete character of any complex dynamics following, in its turn, from the unreduced, holistic nature of any interaction process and its adequate presentation within the dynamic redundance paradigm. Each next discrete 'energy level' (or any other 'eigenvalue') in a quantum system appears due to addition of the same complexity quantum, $h$, expressed in the corresponding form (energy, momentum, etc.), to the actually observed level of interaction process (usually due to complexity exchange between various participating entities) [7,11]. This is the nearest higher level of complex dynamics of the world with respect to that of the 'pure' quantum beat of a 'free' field-particle, where the quantum-beat discreteness is still clearly felt and is naturally and 'coherently' included into the discreteness of interaction between several (or many) field-particles (the latter processes can therefore be considered as 'superstructures' formed over the 'elementary' structure of the 'free' quantum beat process).

The extended wave equations are interpreted as the delocalised, undular form of description of the unified quantum beat process in the system of interacting protofields, dualistically (and thus dynamically) related to the corresponding localised, 'corpuscular' (or 'classical') form, as it was explained above. It is very important also that the causally explained effects of 'special relativity' are naturally included into both these dual forms of description, being another manifestation of the same dynamic complexity of the field-particles and their higher-level interaction, which is quite different from mechanistic, axiomatic 'joining' between quantum theory and relativity in the unitary science. The wave equation for the *e/m waves* (*dynamically* quantized into the photons) naturally follows from the obtained Dirac and Klein-Gordon equations, if we neglect, in accord with the above causal notion of the photonic (e/m) field, the term with the relativistic mass-energy describing the 'bound' protofield oscillations 'within' a massive field-particle [7]. The obtained causal unification of the extended versions of the basically separated domains of the canonical fundamental physics includes also the realistic version of 'particle physics', since we obtain now the clear, physically complete understanding of the entity of 'elementary particle' (exemplified here mainly by the electron, the simplest species), in the form of a regime of the complex-dynamical quantum beat process in the unique system of coupled protofields/media (see [7] for further details).

The last domain of the fundamental physics that needs to be included in the same extension and unification procedure constitutes the central subject of this work: it is the theory of gravitation, or 'general relativity', in terms of a canonical interpretation. As was explained above, the universal phenomenon of gravitational attraction between field-particles and thus any many-particle bodies naturally emerges in the proposed physical system of interacting protofields, or 'world sandwich', as effective, indirect interaction between the particles transmitted through the gravitational protofield and appearing because each particle directly modifies the mechanical properties (tension) of the gravitational medium in course of unceasing series of reduction-extension cycles (quantum beat process). This naturally emerging interpretation of the universal gravitation includes its fundamentally *quantum* origin and character (actually perceivable at the smallest scales of space and time), the causally understood *equivalence* between the inertial aspect of mass-energy, determined as the quantum-beat intensity (temporal rate), and its gravitational aspect characterising the magnitude of the resulting gravitational attraction, and the essential modification of the 'geometric', mechanistic approach of the canonical 'general relativity' (see also [7,11] for more detail). The appearing spatial distribution of tension in the physically real gravitational protofield permanently changes at various scales forming what is usually called 'field of gravity' and can be described by the 'gravitational potential', $\varphi(x,t)$, which acquires thus a physically consistent interpretation (in practical observations it appears as the 'potential energy' of a body in the 'gravity field' divided by the mass of the body). Now if the quantum beat process of a field-particle is currently situated at a point with certain value of the gravitational potential (i. e. gravitational protofield tension) its frequency, determining according to eq. (9) its mass-energy and various other properties, will take the value depending on this tension and thus on the point coordinate (for simplicity we consider the case of quasi-static gravity field, cf. [26]):

$$h\nu_0(x) = m_0 c^2 \sqrt{g_{00}(x)} \ , \tag{29}$$

where the classical 'metric tensor' component, $g_{00}(x)$, should be understood now not as the characteristic of the postulated 'curvature' (or 'geometry') of the abstract (mathematical) 'coordinate manifold', but rather as an expression containing the gravitational potential and taking into account its (generally nonlinear) relation to the quantum beat frequency at that point (for the case of weak fields, $g_{00}(x) = 1 + 2\varphi(x)/c^2$, where $\varphi(x)$ is the classical gravitational field potential [26]). Since $\nu_0(x)$ determines the causal 'time flow' (see above) and $\varphi(x)$ has the negative sign ($g_{00}(x) < 1$), the above equation substantiates the causal extension of the well-known 'relativistic time retardation' in the gravitational field.

We should emphasize the fundamental, qualitative difference of the very type of the emerging irreducibly *dynamical* interpretation of gravity as a *complex-dynamical process* from the invariably mechanistic, clockwork, 'geometrical' and absolutely non-physical interpretations of gravity by either standard 'relativity', or any its 'quantum' and 'field-theoretic' interpretation within the unitary (single-valued) paradigm of the canonical science. As any other phenomenon in the unreduced, adequate picture of the real world, provided by the dynamic redundance paradigm, gravitation and any its particular manifestation is a 'living', permanently internally changing, dynamically multivalued and therefore unpredictable in details (probabilistic) phenomenon. Already the simplest relation of eq. (29), provided with the above picture of the dynamically complex (i. e. *essentially nonlinear*, *chaotic* and *fractally structured*) quantum beat process (see eqs. (6), (7) and the accompanying discussion), clearly demonstrates the *intrinsically quantized* character of *any* gravitational interaction, the property that *cannot* be obtained, or even consistently introduced 'by hand', within the canonical 'geometrical' dogmata of the canonical relativity and various its unitary modifications. The reason for this is clear: the unreduced complex-dynamical character of the quantum beat process would imply that the 'curvilinear coordinates' of 'lines of equal stress' (or 'geodesics') in the formally 'deformed', mathematical space-time should constitute the dynamically *fractal* (and thus, in particular, causally *probabilistic*) network, which can hardly give an efficient description and in any case does not correspond to any real, physical deformation. The formal description in the curvilinear system of coordinates is used in various problems of physics, but does not necessarily imply a real deformation of space or any other physical entity (and in any case an actual *deformation* of something necessarily implies the clear definition of the tangible, physical nature of this 'something', which is not the case for the fundamental, 'embedding' space and time when they are considered in the framework of the canonical, unitary science). In addition, one can consistently describe in this fashion only artificially over-simplified, regular (single-valued) configurations, which is not the case for generic problems in the three-dimensional space usually containing at least several independent (and interacting) objects. It is not surprising therefore that the standard relativity seems to be correct (including its 'agreement with experiment') only for that kind of pathologically simple, effectively one-dimensional, and necessarily *macroscopic*, situations, and in such cases it will *formally* — but not physically — agree with the causally complete, complex-dynamical description of gravity. However, once the omnipresent dynamic complexity of the really occurring 'gravitational' processes starts showing its *externally* irreducible manifestations — and this always happens at the smallest, 'quantum', and largest, 'cosmic' scales — the fundamental insufficiency of the 'curved space-time' becomes explicitly evident and leads to the current deep impasse in the fundamental physics (see also sections 1 and 4). We can only add that the absolutely unphysical 'mixture' of space and time into one 'manifold', already discussed above for the case of 'special relativity', as well as 'deformation' of those *purely abstract* 'coordinates' considered, in particular, at the scale of the *whole* world leave no place at all for any real, physical causality in the standard relativity (in a 'strange' contrast to the vigorous fight of its creator, A. Einstein, for 'causality' in another field of 'new physics', quantum mechanics).[10]

---

[10]Apart from the evident subjective origin of such striking difference between the general criteria of causality applied in different fields of fundamental physics, there is an objective difference between manifestations of unreduced complexity in quantum and relativistic phenomena which shows that the

Although the complete mathematical formulation of the causally extended description of gravity needs further efforts, the physical and mathematical consistency of its already realised parts leave no doubts in the right choice of the approach and direction of development. In addition to the mentioned results, they include, for example, the causally derived version of Dirac equation for electron subjected to action of gravity and e/m fields, or physically consistent interpretation of the most compact states of matter within (causally understood) 'black holes', etc. (the details can be found in the book of ref. [7]). Here we can present a simple modification of the basic energy partition relation, eq. (23), for the field-particle moving in a static gravity field and provided with the same causally complete picture of complex quantum beat dynamics behind it. In accord with the classical theory [26] and its quantum interpretation of eq. (29), the total energy conservation means that

$$E = \frac{m_0 c^2 \sqrt{g_{00}}}{\sqrt{1 - \frac{v^2}{c^2}}} = \frac{h \nu_0(x)}{\sqrt{1 - \frac{v^2}{c^2}}} \quad , \tag{30}$$

where $v = v(x)$ is the local field-particle velocity measured in the proper time, and $g_{00} = g_{00}(x)$ is the 'time' component of the classical metric tensor of the static gravity field, but now interpreted as a function of the local gravitational potential, or tension of the gravitational protofield/medium, $\varphi(x)$ expressing its relation to the local quantum beat frequency (for example, $g_{00}(x) = 1 + 2\varphi(x)/c^2$ for the weak potential case), instead of the canonical 'deformation of the space-time manifold along the temporal dimension'. By analogy to our previous analysis (see eqs. (11), (23)), we can try to decompose the total energy-complexity of eq. (30) into two parts corresponding respectively to the global translational motion and chaotic virtual soliton wandering

---

detailed understanding of causality by A. Einstein and his intrinsic followers in *both* relativity and quantum mechanics corresponds rather to the mechanistic, 'symbolical' world of the effectively one-dimensional thinking, than to the physically real, unreduced, and therefore always dynamically multivalued world. Indeed, the main 'causal' objection of Einstein against quantum mechanics deals with the very *existence* of the irreducible indeterminacy he persistently denied *irrespective* of its origin (the very possibility of a dynamical, and therefore *physically* causal, origin of the *irreducibly* probabilistic phenomena has never been considered within this purely mechanistic, one-dimensional 'causality'). Moreover, the further search by Einstein of a 'really causal', and thus necessarily unified, 'field theory' proceeded invariably in the direction of the 'doubly unitary (!) field', that is the unique entity capable to produce all other entities and phenomena through *essentially single-valued*, and actually purely abstract, 'mathematical' mechanisms in the form of 'exact solutions', etc. One finds the same type of mechanistic 'causality' in his 'relativistic' interpretation of Lorentz-Poincaré relations between time, space and motion, where it is reduced to the *postulated*, and actually rather formal, limitation of the velocity of any real displacement to the velocity of light. Such one-dimensional 'causality' has come to an end now, together with the single-valued paradigm of the canonical science, and should be replaced by the *intrinsically complete causality* of the universal science of complexity which can be formulated as the same unique law of the total dynamic complexity conservation (or 'universal *symmetry* of complexity') that can be realised only by unceasing, autonomous, and internally *continuous* (though very *nonuniform*) *growth* of the observed complexity-entropy at the expense of the hidden complexity-information, both these types of dynamic complexity taking *physically real* forms [7]. This rigorous definition of the extended, physical causality within the adequate knowledge incorporates the demands for physical realism, and natural, autonomous, and thus 'first-principle' (purely dynamical) emergence and behaviour of all existing entities.

around this average tendency, both of them acquiring now the additional coordinate dependence (the result should be rigorously valid for sufficiently smooth field inhomogeneities, similar to the 'semi-classical approximation'):

$$E = h\nu_0(x)\sqrt{1 - \frac{v^2}{c^2}} + \frac{h}{\lambda_B}v = hN(x) + h\nu_B(x) = -L + pv, \quad (31)$$

where

$$L = -hN(x) = -h\nu_0(x)\sqrt{1 - \frac{v^2}{c^2}} = -\frac{(m_0c^2)^2}{E}g_{00}(x) \quad (32)$$

is the Lagrangian characterising, up to the sign, the 'thermal' part of the particle energy (totally irregular internal wandering of the virtual soliton, cf. eq. (25)),

$$\lambda_B(x) = \frac{h\sqrt{1 - \frac{v^2}{c^2}}}{m_0v\sqrt{g_{00}(x)}} = \frac{hc^2}{Ev} = \frac{hc}{\sqrt{E^2 - (m_0c^2)^2 g_{00}(x)}} = \frac{\lambda_C}{\sqrt{\left(\frac{E}{m_0c^2}\right)^2 - g_{00}(x)}} \quad (33)$$

is the de Broglie wavelength ($\lambda_C = h/m_0c$ is the Compton wavelength), and

$$\nu_B(x) = \frac{m_0v^2\sqrt{g_{00}(x)}}{h\sqrt{1 - \frac{v^2}{c^2}}} = \frac{E^2 - (m_0c^2)^2 g_{00}(x)}{Eh} = \frac{E}{h}\left[1 - \left(\frac{h\nu_0(x)}{E}\right)^2\right]. \quad (34)$$

Performing now the transition to the nonrelativistic limit (low velocities, weak fields) in the above expression for Lagrangian, eq. (32), we easily recover the nonrelativistic Lagrangian modified by the rest energy (cf. [26]):

$$L = -m_0c^2 + \frac{m_0v^2}{2} - m_0\varphi(x). \quad (35)$$

The physical sense of the obtained result, eqs. (31)-(33), in terms of the quantum beat dynamics is evident: as the field-particle 'falls down' in the gravity field, globally moving to locations with higher absolutely values of the potential, its regular-motion component grows, which corresponds to a decrease of the de Broglie wavelength, eq. (33), while the purely irregular, 'thermal' part of dynamics, described by the Lagrangian, eq. (32), diminishes leading to relative retardation of the proper time, the latter being just *provided* by those irregular 'jumps'. The fall cannot continue infinitely, since the effective 'vessel' of the field squeezes during the fall and becomes finally too narrow to contain all the irreducible realisations of the system ('thermal motions' of the virtual soliton). The resulting equilibrium distribution of the field-particle density, valid for arbitrary field inhomogeneity, can be found from a wave equation obtained by the causal quantization procedure described above and generalising the Dirac equation [7]. This simple example demonstrates the kind of physically complete, transparent picture obtained in our approach to gravity and naturally including all the 'quantum' effects (without 'mysteries'), causal wave-particle duality, 'time relativity' etc. We shall see in the next section that this holistic, strictly first-principle description of the universal science of complexity leads also to important, large-scale practical conclusions about the elements of the fundamental structure of the world to be expected and the ensuing directions of research in particle physics, cosmology and other involved fields of the basic science.

The obtained causal interpretation of the universal gravitation in the system of coupled protofields, alias the World, closes the general picture of the lowest sublevels of the system complexity including the naturally emerging and intrinsically unified, causally complete extensions of all fundamental phenomena studied within different, and irreducibly separated, approaches of the canonical fundamental physics, such as quantum mechanics, relativity, field theory, particle physics, and cosmology. The universality of the natural complexity development process based on the dynamic redundance paradigm and the ensuing symmetry of (unreduced) complexity is explicitly present in our truly first-principle approach from the beginning (see eq. (1)) and is further confirmed by the analysis of progressive complexity development to its higher levels starting from more complex physical systems and having no fundamental limits at least up to the highest known fields of the canonical knowledge usually classified as the humanities [7]. At each already formed level of complexity, the naturally emerging objects/entities that were causally *obtained* by the described method of effective dynamical functions *without any simplification* of the really occurring, *dynamically multivalued* processes, immediately start 'replaying' fundamentally the same complex-dynamical interaction process leading to a new generation of (moving and interacting) real objects or 'dynamical regimes' (realisations), etc. The basic mechanisms and main lines of description of all the observed phenomena outlined above remain universally applicable at any level of this developing complexity, including generalised 'motion' (an increased complexity state); causally emerging space and time (acquiring now the hierarchical, multi-level structure) and their 'relativity' at *any* level (relation to motion and interaction); autonomous creation of new entities/realisations by the dynamic, fractally structured entanglement of lower level realisations; generalised 'duality' between the localised 'regular' realisations and the extended 'intermediate' realisation relating them by chaotic transitions; the corresponding *universal* Hamilton-Lagrange-Schrödinger formalism having its related local (trajectorial, Lagrange) and nonlocal ($\psi$-functional, Schrödinger) versions, etc.

One of the important lower sublevels of this universal complexity unfolding process was presented above as the level of elementary *classical* (well localised) systems naturally emerging in the form of the simplest bound systems of field-particles (atoms, etc.). Two other lower sublevels from the same 'quantum' group involving generic types of interaction processes between field-particles are worth mentioning here: they are designated as *quantum measurement* and *quantum chaos*. The quantum measurement involves interaction between an 'essentially quantum' system (often the elementary field-particle) called here the 'measured object' and another, also generally *quantum*, system possessing, however, a strong enough connection to the hierarchy of higher levels of complexity and being thus a *dissipative* system (according to the *rigorous* definition of dissipativity within the universal science of complexity) that serves therefore as the 'instrument of measurement' [7]. According to the above analysis applied now at this higher sublevel of complexity [7,9,11], the total system of object and instrument will undergo the process of transient, but *physically real* reduction-squeeze (or 'collapse') to a localised realisation 'chosen' in a *causally random* fashion among their complete, dynamically redundant set with the probability determined by the causally substantiated 'Born's probability rule'. This 'result of measurement' will be 'registered' also at higher, 'classical' (macroscopic) levels of the 'instrument' due to the dissipative amplification in the hierarchy of complexity levels described above. A transient 'classical' type of behaviour appears, however, already at the lowest, *generally quantum* (microscopic), and *decisive*, level of the measurement process due to a short-living *localised* state of the dynamically entangled 'object' and 'instrument' formed in the process of physically real reduction-squeeze, which provides the causally complete meaning for the inevitable involvement of 'classicality' in the quantum measurement process, just empirically postulated in the canonical theory [32]. The same generic interaction process between two quantum systems, but *without* any noticeable dissipativity in any of them, takes the form of (genuine) quantum chaos [7,8,11]. In this case the redundant realisations also appear, in accord with the universal FMDF mechanism presented above, but they will usually have a delocalised, extended (though inhomogeneous) internal structure, so that the generalised, causally probabilistic 'reduction' of the compound system to each of them will appear not as the spatial 'squeeze' to a narrow 'point', but as a system transition to a new dynamical configuration/regime. The general chaotic behaviour of such *Hamiltonian* (effectively non-dissipative) quantum system will have the form of unpredictable, 'spontaneous' change of incompatible dynamic regimes/realisations and the corresponding measured quantities which can often be registered only by their average, or 'expectation', values. Note that such *true* quantum chaos, including explicit and permanent random changes of the observed system properties, is fundamentally different from its *basically wrong* simulation within the corresponding concept of the canonical, *single-valued* approach (see the relevant references in [7,8]) always analysing *only one* realisation from their *really existing* complete set and thus incorrectly rejecting, within its *irreducibly perturbational* approach, the very essence of 'chaos' as such. In particular, contrary to its unitary simulations, the causally complete, truly complex version of quantum chaos naturally passes to its classical version in course of the usual 'semi-classical' transition (the formal limit of $h \to 0$), this 'classical Hamiltonian chaos' *also* being now *consistently* defined as the *fundamentally probabilistic* change of distinct, incompatible realisations, instead of its canonical, single-valued simulation by a quite *regular* property of 'exponentially diverging trajectories', etc. (see [7] for further details). Needless to say, the regimes of true quantum chaos and quantum measurement can readily occur in combination with each other when one measures, for example, the properties of (one realisation of) a Hamiltonian quantum system with the help of the proper 'instrument', etc. All these results, demonstrating numerous *independent* correlations with experiment and among themselves, provide the *unique*, *consistently unified* system of convincing confirmation of the proposed holistic picture of the world dynamics at *all* levels of its unreduced complexity.

# 3. Physically real Planckian units of space, time, and mass and the causally substantiated spectrum of elementary particles

The presented intrinsically unified picture of the dynamic world organisation at its most fundamental levels, based on the naturally emerging dynamic redundance concept, resolves all the major difficulties of the unitary knowledge by providing the physically complete and mathematically consistent explanation for the well-known facts taken into account by the 'mysterious' postulates of the canonical quantum mechanics, relativity, and other basically separated fields of the scholar science. One may argue that any true progress of knowledge should not only consistently explain the known facts, but also provide a number of new, verifiable predictions. In this section, presenting the central emphasis of this work, we describe such practically important predictions of the quantum field mechanics directly based on the obtained qualitatively new picture of the world dynamics and implying an equally new strategy in further fundamental research.

Note, however, that the canonical, single-valued paradigm has now covered practically all the available space of the directly (and largely also indirectly) accessible *empirical* fundamental knowledge, entering thus into a specific 'final' stage of the 'end of science' [7,12]. This means, in particular, that further real development of the fundamental knowledge can be performed within a qualitatively new approach to understanding of those *formally* 'known' (but in reality only positivistically 'registered') empirical facts. Of course, the new facts can, and will, eventually emerge in the expected further development of the new paradigm, but its main feature, especially at the beginning of its emergence, should necessarily consist in a really consistent explanation of the *known* facts efficiently eliminating *all* the existing contradictions. Therefore the *new* predictions described in this section are also largely based on essential *modification* of the existing unitary-science notions which nevertheless can lead, as we shall see, to the dramatic change of the whole system of priorities in science and reorientation of many its fundamental domains. We argue that such *causally substantiated choice of direction of development* (ensuring the optimal *growth* of the unreduced dynamic complexity) should in general be the *main* goal of the *fundamental* science at the forthcoming qualitatively new, really *conscious* stage of its development (coinciding, as we show, with the universal science of complexity), as opposed to the irregular trial-and-error type of single, basically *separated* empirical 'discoveries' of the late canonical science, which are only *externally* (and often incorrectly) 'arranged' in a system of *abstract* symbols with the help of the positivistic criterion of 'coincidence between theory and experiment' (actually always subjected to unlimited mechanistic 'adjustment' with the help of arbitrary 'parameters', 'models', and 'postulates' and therefore providing only a tricky *imitation* of the proclaimed 'practical importance of science').

Since the basic separations of the 'symbolical' notions of the unitary science cannot be eliminated by any efforts within the same paradigm, its adherents try to attain the desired 'unification' of knowledge by purely mechanical 'joining of symbols' which cannot even pretend to be explained by a consistent physical picture. Those cabalistic games taking 'scientifically looking' names of a 'dimensional analysis', 'renormalisation group', or 'Feynman diagrams' (and other abstract 'graphs') lead but to thoroughly adjusted sets of numbers demonstrating, of course, 'coincidence with experiment' by the very method of their production and irrespective of the physical reality they are supposed to describe (e. g. [33,34]). In fact, this kind of 'science' is so 'fundamental' and so 'prodigious' (according to the self-estimate of its 'wunderkind' creators) that it does *not* need to contain any genuine, material reality *at all*: in a characteristic similarity to some 'gross' financial account, *everything* is reduced to correlations within systems of numbers, and the 'naive' physical questions about the material nature of the described entities and detailed, causal mechanisms of their occurring transformations simply have no place within the mechanistic thinking of the modern scientific 'scribes'.[11]

The status of conventional 'Planckian units' belongs to this kind of purely technical 'arrangement' of quantities achieved by the canonical 'dimensional analysis', even though Max Planck, introducing the idea a hundred years ago, supposed probably its forthcoming development towards a more profound, physical foundation. The particularity of the Planckian units is that they 'realise' one of the most ambitious unifications of the modern physics, involving gravitation, electro-magnetism and quantum mechanics, by very simple, and still apparently irreproachable, means, i. e. by combining the Newton gravitational constant, $\gamma$, the speed of light $c$ and the Planck constant $\mathrm{h}$ into the unique possible combinations with the dimensions of length, time, and mass. One is *forced* to assume then, within the 'method' of mechanistic science, that those scales having actually ultimately small (length and time) or big (mass) values should correspond, at least approximately, to some really existing elementary entities (most probably particles) which mark the 'borders' of the world construction. Needless to say, no any physically real picture of actual origin and mechanism of emergence cannot consistently underlie this assumption within the unitary science; such more detailed theory is suggested to be feasible within much more elaborated, but also purely abstract, constructions of the 'unified field theory' proposing actually the absolute minimum of transparency and the maximum of undecidability due to their (non-dynamical!) redundance.

In particular, since the conventional Planckian mass unit, $m_\mathrm{P} = (\mathrm{h}c/\gamma)^{1/2} \approx 2 \times 10^{-5}$ g ~ $10^{19}$ GeV/$c^2$, is extremely high with respect to the heaviest observed elementary particles (it is $10^{17}$ times greater than the highest really observed elementary object mass, of the order of $10^2$ GeV/$c^2$), the experimental 'high-energy physics' is forced to look for new particles by passing to 'super-high' energies attainable only at the super-expensive super-accelerators, but even the largest of them will never be able to span at least a reasonably small part of the huge, now empty energy gap limited from above by $m_P$. Apart from the evident, and actually irresolvable, practical problems involved and the extremely unpleasant general situation of fundamental impasse in this search, there is the equally striking and fundamental contradiction with the very general and always confirmed properties of nature organisation around the 'principle of parsimony' ('Ockham's razor' in theory). Indeed, the mass-energy interval between the lightest and the heaviest particles (i. e. between 0 and $10^2$ GeV) is 'filled up' by various species in a reasonably 'dense' fashion, though maybe inhomogeneous in details (and such nonuniformity is also the necessary property of a realistic structure distribution, acquiring its fundamental substantiation only within the science of complexity). However, outside this 'normal' energy interval there is a strange, and extremely huge, 'emptiness' which *cannot* simply be the end of the elementary world structure (with its border just around $10^2$ GeV), within the canonical approach, because the latter predicts that *some* objects *should* necessarily exist until the mass-energy of $10^{19}$ GeV, or at least its reasonable portion. Despite this evident contradiction with the principle stating that 'nature does not support (big) emptiness', the canonical value of Planck mass is very widely used in both theoretical efforts trying to adjust their parameters so as to obtain this expected 'something' within the gap and related practical planning of experimental search for hypothetical 'new species'. As a result, the modern 'particle physics', 'high-energy physics', 'field theory', cosmology, and the related subfields are helplessly submerged into the steady bog of arbitrary trial-and-error search without issue producing no expected discoveries, but only unlimited number of new *abstract* possibilities always readily 'substantiated' by a deceitful game of fitted numbers and ambiguous symbols.

---

[11]When A. Einstein supposed, within his enthusiastic critics of the canonical interpretation of quantum mechanics, that 'God does not play dice', did he implied, by further constructing his purely mechanistic interpretation of the Lorentz-Poincaré relations between space-time and motion, that the 'God' of his faith performs instead the purely calculative work of accountant? It is not surprising, for example, that the 'problem of ether' as a necessary *material basis* for perturbations known as e/m waves simply 'disappears' from the canonical 'relativity' that does *not* need *any* material, physically real basis for its purely abstract, artificially imposed, and physically senseless 'principles'. But "whatsoever a man soweth, that shall he also reap", and the fraud of superficial mechanistic substitutions of the canonical 'new physics' brings forth its bitter fruits today, in the form of the global crisis of the 'end of science', after which any further progress is strictly impossible unless everything in science is remade from the very beginning and provided now with a *physically* complete, really *causal understanding*.

A similar contradiction exists for the conventional values of Planck length, $l_P = (\gamma \hbar/c^3)^{1/2} \approx 10^{-33}$ cm, and Planck time, $t_P = (\gamma \hbar/c^5)^{1/2} \approx 10^{-43}$ s, which are, contrary to $m_P$, excessively small, but as we shall see, this is because of the *same* fundamental deficiency of the standard mechanistic approach as the huge value of the Planck mass. The unrealistically small values of Planck length and time are equally widely used in various hypothetical constructions of the unitary science, including especially popular ideas about the 'space-time foam' implying that the 'space-time manifold' from the standard relativity is actually structured into fine elementary 'cells' with spatial dimensions of the order of $l_P$ that should (chaotically) 'flicker' with a time period of the order of $t_P$. Since those hypothetical elementary blocks of the space-time structure are 16-19 orders of magnitude smaller than the smallest really observed dimensions of elementary objects, one obtains again a strangely large 'gap of scales' that can reasonably be filled only with unrealistically high number of strangely 'hidden' species. Those guesses about the 'structure of vacuum' are closely related to mechanistic constructions of the canonical 'quantum gravity', but all of them are eventually only *postulated* in the form of pathologically simplified mathematical structures which are fundamentally detached from the physically real entities and processes they are trying to simulate, so that the causal meaning of the very notions of space, time, and mass remains inaccessible, to say nothing about their dynamical structure and its origin. This leads inevitably to the same arbitrarily changing, basically indefinite choice of abstract 'models' only deepening the fundamental impasse of the one-dimensional thinking.

It is quite natural that the approach of the universal science of complexity, proposing the first-principle, progressive derivation of the physically real entities in exact correspondence with their actual emergence, provides the fundamentally substantiated, basically simple and physically transparent resolution of the mentioned difficulties of the unitary science involving a well-specified physical meaning of the Planckian units and the ensuing essential modification of their values, which should have, as we shall see, important practical consequences for the related fields of theoretical and experimental research providing it with a fundamentally substantiated general criterion and direction.

The physical, causally complete origin of 'Planckian' type of sizes, combining gravitational and e/m degrees of freedom and representing the extreme characteristics of elementary objects, immediately follows from our unified picture of complex world dynamics at the level of quantum beat processes within each elementary particle. Namely, it is clear that the (probably modified) Planck length corresponds to the (smallest) size of the 'virtual soliton', that is the elementary field-particle in the phase of maximum reduction-squeeze, the Planck time describes the time period of the reduction-extension cycle of the most rapid quantum beat process (for the same field-particle), and Planck mass is the (largest observable) complex-dynamical mass-energy of the same particle-process, all those quantities naturally attaining their respective extreme observable values for that particular type of the elementary world structures. We emphasize the well-specified physical meaning of the extreme quantities thus introduced: they cannot be separated from particular physical structures of the type of *well-defined*, 'individually' appearing object-process-particle and *cannot* be associated with only a vaguely conceived 'background structures' like a quasi-homogeneous 'foam', 'structure of vacuum' ('virtual particles'), or any other 'noise'. Indeed, already the *causally specified* property of mass inevitably involves some chaotic 'virtual soliton wandering', and thus a 'reduction-extension process', and therefore the squeezed, corpuscular phase of that process, etc. The same refers to the causal concept of time: a meaningful time 'interval' can only be really, physically 'produced' by the elementary, but well-defined, non-linear, 'structure-forming' (squeezing) type of 'pendulum'-oscillation, whereas a random 'flicker' of a distributed, ill-structured 'froth' would produce an ill-defined, inhomogeneously varying 'embedding' time.

Now, recalling our causal interpretation of the universal gravitation, one can easily see that this physically consistent understanding of the sense and origin of Planckian scales enters in contradiction with their conventional expressions obtained by mechanical application of the canonical 'dimensional analysis'. Indeed, the Newtonian gravitational constant $\gamma$ entering those expressions characterises in reality the *indirect*, 'long-range' interaction between *two (or many)* individual field-particles (or many-particle bodies) transmitted through the (hidden) gravitational medium due to the unceasing reduction-extension (quantum-beat) processes within each particle, whereas the physics of Planckian units is determined evidently by the parameters of the fundamental *direct* interaction-coupling between the e/m and gravitational media, giving rise to the *quantum beat process itself* within *each single* field-particle. It is clear also that the characteristic magnitudes of those two interactions, expressed by the respective 'constants' and incorrectly confused in the purely formal 'dimensional analysis', *should* be quite different: whereas the observed, *indirect* gravitational interaction is much *weaker* than the more direct e/m interaction between the field-particles (transmitted through the perceivable, e/m side of the 'world sandwich'), the *direct* coupling between the protofields needs to be much *stronger* than the ordinary e/m interactions, since each of the field-particles should preserve its characteristic reduction-extension dynamics despite any 'usual' inter-particle interactions. Therefore, the effective constant of the direct protofield coupling, $\gamma_0$, that should *actually* enter the genuine, *causally corrected* expressions of the Planckian units instead of the ordinary gravitational constant $\gamma$ used in their canonical form, is much greater than $\gamma$, $\gamma_0 \gg \gamma$ (whereas the algebraic structure of the expressions cannot change, as we confirm below). This will lead to changes of the Planckian scales just in the 'good' direction permitting us to avoid the described contradictions of the strangely extreme, unrealistic values from the canonical analysis and at the same time confirming the proposed physical interpretation of these fundamental units: the space and time units will *grow* to their *actually observed* extreme values for the corresponding elementary field-particles, while the mass unit will drop also to the actually observed largest mass of the same 'extreme', but *real* field-particles. Taking into account the big amount of the accumulated experimental data and also the evident sufficiency (and even some well-known redundancy) of the discovered particle species for the viable world construction (cf. the 'principle of parsimony' of nature), one can be quite sure that the real values of the Planckian units thus obtained and interpreted correspond indeed to the extreme *already* observed characteristics of the elementary objects (the deviations of one-two orders of magnitude cannot be excluded, but they are not essential in this context). If we designate the *modified*, true values of the Planck quantities by the respective upper-case letters, $L_P$, $T_P$, and $M_P$, then the above conclusions can be summarised as

$$L_P = (\gamma_0 h/c^3)^{1/2} \approx 10^{-17}\text{-}10^{-16} \text{ cm},  \tag{36}$$

$$T_P = (\gamma_0 h/c^5)^{1/2} \approx 10^{-27}\text{-}10^{-26} \text{ s},  \tag{37}$$

$$M_P = (hc/\gamma_0)^{1/2} \approx 10^{-22}\text{-}10^{-21} \text{ g},  \tag{38}$$

which shows, by comparison to the conventional values, that $\gamma_0 = 10^{33}\text{-}10^{34}\gamma$. Among the values on the right, the value of (largest) experimentally observed mass of elementary particle (of the order of 100 GeV, as that for intermediate bosons $W^\pm$, $Z^0$) seems to be better established, and we actually referred the value of $\gamma_0$ to this quantity, while calculating the units of length and time for this value of $\gamma_0$. Note, however, the remarkable agreement of the latter values with the smallest observed (or just experimentally 'indicated') sizes and time periods, taking into account larger difficulties in experimental approach to the ultimate existing scales of these quantities.

Not only the obtained results provide the physically consistent, causally complete picture of the elementary building blocks of nature, but they immediately lead to important practical consequences for the whole strategy of experimental research in the field. It becomes clear that one does not need to attain some 'super-high', and actually unrealistically high, energies in accelerator collision experiments in order to span the *whole* mass spectrum of the existing elementary species: one would need at maximum the effective collision energies exceeding $M_P c^2$ by several orders of magnitude, and since this is already realised at the existing facilities, the false necessity to deal with $m_P c^2 = 10^{19}$ GeV from the canonical, unitary approach remain only a nightmare of the one-dimensional thinking (by its over-simplified concept and over-sophisticated technical constructions the single-valued science has become a nightmare even in its purely theoretical aspect!). Further research should be concentrated on the details — now also progressively *re-interpreted* within the *causally complete* approach of the quantum field mechanics — hidden *within the known spectrum of mass and interactions* and be directed towards (now quite meaningful and feasible) *ultimate completion* of the *unreduced* (complex-dynamical) physical picture of the fundamental levels of reality. The realisation of this task will certainly entrain a further series of various practically important applications — eventually the *largest possible* scope of applications (since they are based on the *physically complete* understanding of reality) — and should, in principle, even show the way *beyond* the limits of *this*, 'our' world construction (the 'world sandwich' of the two coupled protofields) which are certainly rather 'rigid', but *finite* and *cannot* be absolute or 'irrational', as it is *rigorously* shown within the universal science of complexity [7].

We can further specify the proposed causal interpretation of the Planckian units by elementary estimates directly related to the real physical processes involved. Consider the 'potential energy', $E_P$, of the corpuscular state of the heaviest field-particle with certain mass $M_P$ dynamically squeezed down to the size $L_P$ in this phase of the quantum beat process. Since at those ultimately small spatial scales the difference between the 'direct' protofield interaction and the resulting 'indirect' gravitational interaction disappears, this energy can be estimated as the ordinary energy of gravitational interaction, but where the ordinary gravitational constant $\gamma$ is replaced by its extreme short-range limit, $\gamma_0$, characterising the direct protofield interaction: $E_P = \gamma_0 M_P^2 / L_P$. From the other hand, this energy should be of the order of the total energy liberated in a cycle of the quantum beat, $M_P c^2$, that is

$$M_P c^2 = \frac{\gamma_0 M_P^2}{L_P} \ . \tag{39a}$$

The time period of the quantum beat, $T_P$, is related to the particle mass by the causally completed version of 'de Broglian' mass-frequency relation of eq. (9):

$$M_P c^2 = \frac{h}{T_P} \ . \tag{39b}$$

The third relation between the ultimate scales comes simply from the fact that we deal with the physically real e/m protofield, where all perturbations propagate with the speed of light:

$$L_P = cT_P . \qquad (39c)$$

It is easy to find now that the unique solution of the obtained system of equations, eqs. (39), is given by eqs. (36)-(38), where only h should be replaced by $h$ (this seems indeed to be a pertinent modification that does not change the essential result, however). In this way we confirm the above physical interpretation of Planckian units and the ensuing practical conclusions. One aspect of this interpretation is explicitly involved in the above demonstration and confirms also the whole fundamental construction of the world in the form of the unique couple of the interacting protofield. Namely, the modified Planckian mass-energy, $E_P = M_P c^2$, estimates the {maximum) magnitude of the fundamental protofield coupling itself actually realised in the regime of the most intensive (rapid) complex-dynamical interaction process (quantum beat within an elementary field-particle), in the form of its mass (temporal rate). This interpretation is directly related also to the intrinsic 'unification of particle interactions' (and particle species) realised within our approach from the very beginning in the form of the *dynamically multivalued* interaction process in the *single*, *physically* unified, real, and simplest possible system of two protofields. The obtained result demonstrates the basic property of the basically unlimited, but *unified diversity* constituting the eternal 'super-goal' of any knowledge and uniquely attained within the universal science of complexity due to the causally complete, unreduced character of the dynamic redundance concept [7].

Note also that the modified Planck units of space and time causally interpreted within the quantum beat picture of the elementary field-particle should be considered also as further refinement of our causally complete understanding of the physical origin of space and time (section 2). The (modified) Planck length $L_P$ gives the smallest possible size of the space structure element within our world probably realised as the corpuscular state ('virtual soliton') for the elementary field-particles with moderate masses (like the electron), but also as the 'Compton wavelength' (the characteristic size of the causally random wave field equal to the size of elementary jump of the virtual soliton) of the elementary particle(s) with the mass close to $M_P$: $L_P = h/(M_P c)$. The (modified) Planck time determines the highest quantum beat frequency and thus the unceasing, irreversible flow of the most fundamental level of the hierarchy of time (section 2) in this world. Any interval, $T$, of this most fundamental, 'embedding' time is given by $T = nT_P$, with the corresponding integer $n$. Note, however, that often the role of 'embedding' time unit can belong to the corresponding causal time period emerging in another, less rapid, but actually most important quantum beat process, for example, that of the electron, as far as the levels of electronic, photonic, atomic (and higher-level) phenomena are involved. One can consider that each fundamental type of those 'non-extreme' elementary particles, representing a particular regime of the complex-dynamical process (quantum beat) of the protofield interaction, is characterised by its proper, effective value of the 'protofield interaction constant' $\gamma_0$ giving the respective values of the 'Planckian units' in accord to eqs. (36)-(38) which are, however, less fundamental than the basic units from eqs. (36)-(38), correspond to only a *partial* transformation of the protofield interaction into mass-energy, and determine higher sublevels of the 'embedding' space structure (these details will not normally appear at the level of 'ordinary', higher-level observations, starting already from atomic phenomena).

An additional, less expected application of the causally modified elementary unit of mass $M_P$ can be found in nuclear physics, if we note that the atomic nuclei can be considered, due to the very close, 'mixing' interaction between the nucleons, as 'big elementary particles', though possessing a compound internal structure. The latter factor means that the observed mass-energy of a nucleus can contain some non-negligible contributions from the 'coupling energy' between the nucleons. However, if we deal with rough estimates of mass by order of magnitude, those details cannot influence the conclusion that states that the maximum possible mass of a reasonably stable nucleus should also be of the order of $M_P$. This result directly follows from the very essence of the causal interpretation of mass-energy in the quantum field mechanics as the temporal rate of the (spatially chaotic) reduction-extension events of the *universal* protofield interaction process: whatever is the particular kind of the 'elementary' object, its highest possible mass is always obtained in the regime of *maximum* transformation of the magnitude of the fundamental protofield interaction into the mass-energy of particular complex dynamics of the object which is thus given by the universal value of the fundamental protofield coupling, $M_P c^2$ (taking into account that the object 'elementary' enough, like a nucleus, possesses a sufficiently homogeneous complex dynamics not dissociated into very different levels). Since the actual value of the modified Planck mass, $10^2$ GeV, correlates indeed well, by the order of magnitude, with the masses of the heaviest nuclei, we can say that this fact provides additional, causally substantiated support for both the causally modified interpretation of the Planckian units and the whole complex-dynamical construction of the world in the form of the unique couple of interacting protofields.

The remarkable fact that the characteristics of the spectrum of chemical elements, determining the properties of the directly perceivable, 'macroscopic' and 'everyday' reality, and the spectrum of much more exotic, but fundamentally important objects like elementary particles both originate, now also in a *direct* sense, from the same, naturally unique complex-dynamical process of the protofield interaction and therefore can be causally, physically understood as *emergent* manifestations of this unified world dynamics has further implications for causal understanding of details of the fundamental world structure. Thus, a general type of comparison between the periodic table of elements and the system of elementary particles has been performed, from various aspects, already within the canonical, mechanistic constructions. It is clear, however, that the structure of the table of elements is understood much better (at least at *its* level of complexity), than the system of the elementary particles, which leaves little doubt that the table of elements is reasonably *complete*, whereas the system of elementary particles is always subject to essential modifications and the basic understanding of even its empirically well known parts can never approach the same degree of genuine, physical causality within the canonical, mechanistic 'classification'. The results described in this work show that the fundamental, irreducible ambiguity of the canonical microphysics can be totally eliminated at the expense of accepting the extended and internally complete approach of the dynamic redundance concept, which should result in the same detailed, physically transparent interpretation of all existing fundamental objects as that of the level of chemical elements (many rules of that higher level are also formally accepted from the lower levels of microphysics where they are simply postulated, but can now be *causally* substantiated within the same first-principle picture of the quantum field mechanics).

In particular, we have seen that already within a rather brief analysis presented in this work one can consistently explain the complete physical essence of an elementary particle exemplified by the electron and reduced to a regime of the unreduced complex dynamics of reduction-extension (quantum beat) process in the simplest system of two interacting protofields. And although we cannot observe the detailed picture of the protofield dynamics within such particle-process, the *natural* emergence in this system of such specific properties like wave-particle duality that could not be understood at all within the conventional, single-valued approach provides a number of convincing and independent confirmations for the obtained description. The same causal completeness is preserved at higher sublevels of dynamic complexity with the *emerging* long-range (e/m and gravitational) and short-range (weak and strong) interactions between the field-particles always causally *transmitted* through the *physically real* media/protofields and realising the intrinsic, dynamical 'Grand Unification' of all interactions (see also section 2). The related elementary particle species also naturally and progressively emerge in the same naturally unified process of development of the fundamental protofield interaction. Although we cannot present, within this draft picture, the detailed causal interpretation of all the species and modes of their behaviour, the results already obtained (see also ref. [7]) present a large enough system of independent correlations not involving *any* formally imposed postulate at all for considering creation of the detailed picture of the causally complete world dynamics as a quite feasible task. It is this, and only this, way of development of the fundamental knowledge that can provide also many practically important applications similar to those described in this section.

# 4. Conclusion: Science after the end of science and the urgency of transition to the causally complete knowledge

After having presented, in section 1, our substantiation of the fundamental origin of the modern crisis in science as being due to its basically reduced, single-valued vision of the irreducibly multivalued real world and specified, in sections 2 and 3, the issue from that situation in the form of the universal science of complexity and its application to the lowest levels of the world dynamics, we can now state that the transition to such qualitatively new, and actually the only adequate, form of knowledge is both inevitable for any further progress in science and urgent if we want to really attain this progress and avoid the forthcoming serious problems in the whole civilisation development.

The fundamental, evident end of the canonical, unitary (single-valued) type of knowledge acquires now its consistent, and even in a sense promising, interpretation: the development of science has stopped now because it could never *really* begin within the single-valued (unitary) concept absolutely dominating until now in all fields of properly 'scientific', 'conscious' (not purely empirical) knowledge. Indeed, the actual degree of consciousness in the unitary science does not provide any basic distinction from the purely empirical observation of reality: the proposed positivistic 'ordering' of observations is based *itself* on *empirically* established 'principles' *not* supported by any profound, internally complete understanding. Even so, the very limited progress of true understanding was always possible only due to occasional deviations from the purely mechanistic thinking, though quickly suppressed by the dominating one-dimensional thinking. There were only a few known larger deviations from the reigning mechanism within the modern epoch of knowledge, such as that of the 'universal science' of René Descartes proposed at the very beginning of the 'scientific' knowledge or else the irreducibly causal approach of Louis de Broglie in the recent epoch of 'new physics'. However, many of the well-known results of the unitary science, such as the unified theory of e/m phenomena created by James Clerk Maxwell, appeared originally in a much more profound form tending to physical understanding and only later were schematised by the invariably winning mechanistic thinking.

It is not surprising either that the decisive fall of the basically insufficient single-valued simulation of the multivalued reality happens only now, after several centuries of its externally 'successful' development. This is because among the two characteristic regimes of the unreduced dynamic complexity, the 'uniform chaos' and the (generalised) 'self-organised criticality' (or 'extended self-organisation') [7], only the former presents the explicitly observed, undeniable signs of dynamic multivaluedness in the form of sufficiently rapid change of sufficiently different regimes (realisations), whereas the latter contains the same unceasing realisation change process in a largely hidden, 'well packed' form, and therefore provides a convenient illusion of 'success' of the mechanistic adjustment of an 'exact', effectively one-dimensional solution to the externally observed, rather rigid 'envelope' of the internal 'true life' of the system (in particular, this is the fundamental simplification of the canonical concept of 'synergetics', or 'self-organisation', see [7]). The observed irreducible manifestations of the dynamic randomness were always explained by 'uncontrollable' influences of the 'environment' (the basic inconsistency of such explanation is evident, but it is difficult to rigorously disprove it in practice). The intensive emergence of the 'new physics', starting from the end of the 19th century, had marked the inevitable, growing manifestation of the explicitly complex, causally random (chaotic) phenomena, even though they were not recognised as such, despite some singular efforts of intrinsically causal thinkers, like Poincaré and de Broglie, and the single-valued approach succeeded again in imposing its pathologically simplified constructions for another hundred years. However, since the explicit manifestations of unreduced complexity have now decisively appeared on the scene, this time the price for substitution was much higher than before: the false 'victory' of the one-dimensional thinking now could not avoid the explicit introduction a number of not only unexplained (as usually), but also *very* strange (and *fundamentally* important) postulates and even directly acknowledged 'barriers of cognition', such as the well known 'quantum mysteries' or 'relativistic principles'. The whole twentieth century has been wasted then for useless manipulations around the same false maxima, dogmatically promoted by the same interested 'false prophets'. The result is the practically total absence of any substantial progress in fundamental science during the century naturally ending with the current 'definite' crisis of the 'end of science', which demonstrates a true cost of deviation from the principles of genuine understanding and causality, especially impressive because the chances to ascend to the truly conscious, intrinsically complete knowledge were much more real this time, but were disgracefully and stupidly sacrificed in favour of vain personal ambitions and self-interested promotions. The time spent in thorough exclusion of any attempt of a more adequate, causal understanding of the unreduced, 'living' reality has led to extra-ordinary increase of the effective barrier separating the civilisation submerged into the bog of low-level, mercenary manipulations from the extended, Renaissance type of thinking that can uniquely provide the true, internally complete mastery of the world. The results of the universal science of complexity already obtained show unambiguously that the transition from the single-valued to the adequate, dynamically multivalued description means that everything in science should be redone at the qualitatively new level, which is especially difficult not because of the intrinsic sophistication of the new knowledge (which actually *clarifies* the picture of the world), but because of the accumulated power of massive and 'developed' mediocrity and almost total absence of subjects of *serious* higher-level search based on the *unreduced* complexity *growth* (which is quite different from the proliferating superficial esoterism and dark-side obscurantism of various quasi-intellectual 'sects' and vain 'wunderkinds' in science and beyond).

    Note that this sad, but as we can see now, inevitable situation in fundamental knowledge should not be confused with the accompanying successes in purely technical development (related to the 'applied science'). This has been possible because the type of technology still dominating within the modern type of development and severely limiting its possibilities deals almost exclusively with the well-defined 'self-organised' type of regime, where the inevitable appearance of the basically irreducible

multivaluedness has the form of a 'technical fault' that should (and often can) be eliminated, or 'repaired'. The problems with this 'one-dimensional' type of practical creation are rather well known (modern ecological and psychological difficulties are among them) and will also quickly grow until the unavoidable 'end of (unitary) technology', quite similar to the 'end of science', but arriving somewhat later. Despite this relation, the unitary fundamental science has never contributed as much as it states to the (unitary) technology. Analysing the details of the technical achievements, it is easy to see that they happen largely independent from the false 'development' of mechanistic 'science' as such, by a purely empirical search, though often equipped with the same sophisticated *technical* means as the scientific search.[12] The largest contributions ever obtained by technology from (fundamental) science concern purely *qualitative* information about a fundamentally new type of phenomena providing a new *direction* of technical development, and not at all any formal elaboration (one-dimensional, 'exact' solution) of the mechanistic 'science' occupying the absolute majority (today actually the totality) of its efforts. Even the vanishingly small amounts of useful discoveries preserve their basically *empirical* character within the canonical science, and therefore can well be obtained within its 'applied' branch directly attached to the respective practical domain and thus more efficient.

   The always growing crisis in the fundamental science constitutes in itself the reason for searching for an issue, but we have also demonstrated that the proposed well specified, and indeed qualitatively new, approach of the dynamic redundance paradigm *actually* provides the consistent and universal issue from all the known difficulties, beginning from the most fundamental levels of complexity, continuing to its arbitrary levels, and attaining the highest accessible levels (see also [7,8-11]). As shown in section 3, the obtained results lead also to important practical consequences that can, most probably, be considerably developed in subsequent research and are actually supported by the already realised, equally significant applications of the same concept at higher complexity sublevels, such as the (genuine) quantum chaos [7,8] and complex-dynamical extension of 'quantum computing' [7]. Based on these achievements, we argue now that the transition to the proposed new type of intrinsically unified knowledge, referred to as 'the universal science of complexity' [7], is not only desirable, but *urgently* needed. This urgency has two related aspects which we designate as 'conceptual' (theoretical) and 'practical', respectively.

---

[12]Only one particular, but practically important, example concerns the recent development of 'radiation physics' supposed to be 'indispensable' for creation and improvement of nuclear energy sources. Without going into the details of the experience in the field, let us note only that its fundamental part has been formally very well 'developed' in both Soviet Union and Western countries (including the privileged financial support and the obtained indeed large number of 'very interesting' results), but whereas the Western nuclear facilities work quite well, *irrespective* of the fundamental radiation physics, the Soviet analogues created big problems (Chernobyl, etc.), *also irrespective* from 'fundamentals', and could be partially ameliorated only after *purely technical* Western help. From the other hand, the extremely high support of the very ambitious programs on the thermo-nuclear power sources did not bring any of the widely announced results at any national or international scale simply because it is one of the cases where one deals with explicit manifestations of the unreduced complexity fundamentally inaccessible for the effectively one-dimensional, perturbative approach of the canonical science (see also [7]).

The conceptual aspect of urgency reflects the growing bad consequences of the 'crisis in science' as such, involving all kind of internal decay of the system of knowledge acquisition, but especially the rapid degradation of the notions around the 'criteria of truth' and their practical realisation. It becomes today so evident that the canonical science cannot provide a true solution for any of the existing fundamental (and thus the related practical) problems that it is very difficult to avoid serious doubts in the scientific method in general, in any its version (e. g. [5]). While the qualitatively extended approach of the universal science of complexity provides the desired issue from the crisis and way for the new development, it also demonstrates in a much more convincing, fundamentally substantiated fashion the severe limitations of the unitary science, including their origin and the evident insufficiency of the key property of single-valuedness.

In particular, our unreduced consideration of any interaction process in a quite general form applicable to any kind of system (section 2) provides the convincing evidence for the omnipresent multivaluedness which is both rigorous and intuitively transparent. It is quite evident that a *generic*, unreduced interaction between $N$ 'degrees of freedom', where by definition 'everything interacts with everything', should produce a much larger number, of the order of $N^2$, of their combinations forming the new 'degrees of freedom' obtained thus by the *physical* entanglement of the initial entities, whereas the canonical analysis, always using one or another version of basically restricted perturbational approach, gives the result in the form of the *mechanistic sum* (superposition) of the initial entities preserving the same their total number and not involving any *essentially* nonlinear, *dynamical* entanglement. As a result, the unreduced interaction analysis directly predicts the *redundance* of the *incompatible* versions of the forming degrees of freedom (or dynamic regimes) and their ensuing *causally random* 'rotation' (permanent replacement), whereas the canonical science is always reduced to a mechanical combination of *superposable*, effectively *one-dimensional* regimes, including all the *externally* 'nonlinear' *exact* solutions, etc.[13] Needless to say, already a general estimate of actual observations at various levels of reality easily confirms the vision of permanently and unpredictably changing (incompatible) regimes, complementary (or 'dual') versions of behaviour, etc. Despite the evident theoretical blunder, the mathematical inconsistency of perturbational approach (often readily acknowledged), and the omnipresent, persisting divergence with results of observations provoking the current fundamental impasse and long-lasting stagnation, the single-valued paradigm continues to impose its outdated, vulgar simplicity and thoroughly eliminate any signs of extended, dynamically multivalued thinking (thus, the single-valuedness of solutions is always preserved as one of the 'irreproachable', 'main' postulates of the conventional quantum mechanics helplessly trying at the same time to resolve its 'mysteries', including the well-known 'wave-particle *duality*'). This shows that a *conceptual* novelty typically meets much more difficulties, than any purely technical sophistication. The medieval system of 'official' scientific publications using the totalitarian power of well-organised clans of 'referees' in order to eliminate any deviation from the accepted 'absence of anything new' (considered as an indispensable criterion of 'high quality', of course), readily publishes at the same time explicit acknowledgement by the 'officially recognised' science of the too evident fundamental contradictions and insufficiency of its 'solutions' for any consistent understanding of the analysed (in reality always irreducibly complex) phenomena. This situation clearly shows that the existing external 'freedom of speech' and 'democratic' voting for truth not only cannot provide any guarantee against any most evident substitution and fraud, but on the contrary give the maximum freedom for suppression of any genuine progress by all means hypocritically hidden behind the formally 'irreproachable' envelope of (imitated, but actually non-existing) 'equal opportunities'.

---

[13]This concerns also the simulative, actually unitary 'science of complexity' recently appeared *within the same single-valued paradigm* and pompously supported in various and now numerous 'centres for study of complex phenomena' (cf. [12]). In particular, certain substitutions for the genuine dynamic multivaluedness occasionally appearing in that kind of 'advanced unitarity' reveal invariably their empirical character, non-dynamical origin, and narrow limitation to special types of system (called 'complex', or 'multistable', and appearing to be qualitatively different from both the 'ordinary', 'non-complex', generic systems and other types of special, 'complex' systems). As a result, the causal, purely dynamic, and thus omnipresent mechanism of randomness in nature cannot be deduced from the simulative 'multivaluedness' in that kind of studies, to say nothing of the related more involved properties of the unreduced complexity, like dynamic entanglement, generalised 'Born's rule of probabilities', 'intermediate realisation' characterised by the causally extended, universal and physically real 'wavefunction' bonding the multivalued system into a single whole, etc. (the major faults of the single-valued simulation of complexity are further described in ref. [7]).

These general conclusions are confirmed by elementary analysis of critical situation in particular fields of the fundamental physics and comparison to the extensions proposed by the unreduced complexity. Thus, as we noted above, the 'mysterious' quantum *indeterminacy* and *duality* provide even intuitively evident examples of the unreduced redundance that can only have *dynamic* origin at that *fundamental* level of reality (and taking into account the *universality* of its manifestations). The necessity to naturally *obtain* a *localised* state of *physically real* particle clearly points to involvement of *essential* and *universally* defined *nonlinearity* that actually can only take the form of *dynamic entanglement* of the *interacting* physically real (material) fields leading to their autonomous squeeze-reduction (necessarily followed by the reverse extension), as opposed to its unitary simulations by various versions of *abstract* 'state vector reduction' in the canonical, mechanistic science which are actually artificially introduced (postulated) with the help of the ambiguous 'circular' manipulations around the 'density matrix formalism', etc. [7]. The evident inconsistency of the purely speculative, empirical and irreducibly non-universal 'ideas' of the mechanistic science about the origin of 'quantum enigma' is proportional to the number of publications presenting them in the same most prestigious and 'solid' journals that refuse to accept a physically consistent solution. The long list of the vain 'quantum exercises' of the single-valued science reveals but the grotesque game of words deprived from the elementary sense, such as 'decoherence' of 'state vectors' by arbitrary environmental influences 'occurring' within or without some *abstract* 'histories', the artificially conceived involvement in *physical* processes of a purely *mathematical* 'information' or special, 'quantum', form of *abstract* logic, or even the whole 'quantum philosophy' (as if by passing to each new kind of phenomena nature should automatically 'switch' from one type of abstract logic or philosophy to another, simply in order to help some self-'chosen' sages to justify their salaries), the 'orchestrated reduction of quantum coherence in brain microtubules' *directly* involved with consciousness and gravitation simply because all those phenomena are mathematically 'non-computable', etc. The 'force' of such 'strong' theoretic predictions should, of course, be confirmed by the corresponding experimental versions of the same intellectual trickery which tend to arbitrary mix e/m waves, massive particles, their interactions and involve e. g. 'quantum teleportation', 'quantum cloning' (or 'noncloning'), 'quantum information processing', a long series of 'medical' cases of 'quantum science' such as quantum 'tomography' and 'endoscopy', etc. As if it is not enough, the degree of deviation from any realism and causality follows a dramatic increase as one moves from pure quantum mechanics to 'quantum gravity' and 'field theory' filled up with exclusively abstract and increasingly esoteric entities, rules, and symbols, such as 'cellular automata', 'quantum' or 'noncommutative' geometries and topologies, 'graphs' or 'networks' of 'quantum spins', 'strings' and 'M-brains', etc. In a strange contradiction to their fervent belief in abstract symbols as the basis of nature, the adherents of the unitary cabalism request (and obtain) the highest support in quite real, material values, contrary to disappearing defenders of the unreduced realism. In the name of justice one should acknowledge, however, that this modern 'folie' at the border of mental deviation has the 'well established' and very 'solid' precursors appearing especially within the 'new physics' at the beginning of twentieth century (see section 1), such as the canonical 'relativity' whose widely recognised, exemplary 'perfection' within the unitary science is marked by such 'achievements' as arbitrarily imposed postulates about not even symbolical analogues of real entities, but about abstract 'laws of nature' that do not need any specific prototypes in tangible reality at all, or else a *real* deformation of an *abstract* 'manifold' of space ambiguously 'mixed' with time, etc.[14]

---

[14]Moreover, the intrinsically related 'method' of '(democratic) vote for the truth' permitting to the *well-organised* adherents of 'symbolical science' to *dominate practically* despite the *evident* inconsistency of their mechanistic interpretations was also elaborated and 'successfully' tested within the same global movement of 'liberation' towards the 'new physics'. The characteristic examples are provided by the well-known 'Solvay congresses' conveyed starting just from the beginning of the century as if in order to 'clarify' the situation in fundamental physics, but in reality in order to realise a well-organised 'coup d'état' against the irreducible realism of the 'old physics' using, probably for the first time at that scale, the method of informal voting for truth. Indeed, once 'established' by such a 'solid' and 'democratic' discussion, the truth should then be accepted by everybody who ever wants to remain *practically* in science, receive financial help, attention from colleagues, etc. (the dramatic scientific destiny of Louis de Broglie *after* the 'democratic decisions' of Solvay congresses provides a pertinent example, see ref. [35]). In case of the first Solvay congresses, such 'technique', combined with further methods of 'self-organisation' of intrinsic adherents of the one-dimensional thinking permitted them to dominate *even despite* the clearly expressed disagreement from the part of the *original discoverers* of the main effects of the 'new physics', including such really great and renowned thinkers as Lorentz, Poincaré, de Broglie, and Schrödinger (cf. [35]). After this 'successful testing', the system of voting has been technically elaborated, standardised, and transfered to the 'massive production' of science within the modern huge national and international 'research centres', 'science foundations', 'commissariats', 'academies of sciences', etc. As a result, the 'voters for truth' were capable only to propose infinite series of *evidently* inconsistent *interpretations* of the great discoveries, *artificially imposed* as equally 'great achievements' with the help of 'democratic voting', but in really do not containing, since the 'explosion' of the beginning of the century, no any essential advance of the truly fundamental knowledge-understanding. This is not surprising taking into account that a qualitatively new idea is always born in a single individual mind and can at the beginning be accepted only by a small number of others, so that any

One can compare this type of 'causality' with intrinsically, inseparably unified interpretations of quantization, gravitation, 'relativistic' dependence of physically specified time and space on motion, and emergence of elementary particles with their mass and other physically understood properties, all of them being obtained within the dynamic redundance concept as natural and universal manifestations of the same well-defined dynamic complexity (section 2). It is not surprising that the mystification of the canonical 'relativity', inevitably present behind the polished façade of conventional 'postulates', is replaced now by consistent and physically transparent understanding directly related to the 'common sense', so much disputed by the 'prodigious' adherents of the 'new physics'. Thus, the dependence of the causally emerging time on motion and its permanently 'flowing' character are explained by the fact that time is directly *'produced'* by this motion itself as the unceasing sequence of essentially nonlinear reduction-extension events hidden within any type of dynamics. The flow of time is naturally irreversible because the spatial distribution of reduction centres is chaotic (dynamically random) due to the dynamic multivaluedness in the choice of each of them. Any mass changes the average mechanical tension of the physically real 'gravitational protofield' in its vicinity because of the reduction-extension (quantum beat) process, which leads to modification of the reduction-extension frequency for any other massive particle, providing the universal causal mechanism for both 'gravitational attraction' and the effect of 'time retardation in gravitational field'. It is important that the naturally emerging property of universal gravitation is described as an *unceasing* complex-dynamical *process* thus inevitably involving causal quantization and dynamic unpredictability. It is therefore physically (and not only mathematically) 'non-computable', i. e. contains the true, irreducible randomness that can come, by definition, *only* from the *purely dynamic* origin and is *indeed obtained* in this form that does not contain any esoteric *direct* participation of 'conscious thought' (cf. [36]) actually belonging to *much higher* levels of the same complexity. The notion of classical 'ether' also recovers its causal meaning as the necessary material basis for the observed (undular and corpuscular) perturbations. The old 'difficulties' of the classical unitary science around ether are effectively eliminated in its extended, complex-dynamical version in the form of two coupled 'protofields': the unperturbed state of 'free' (noninteracting) protofields cannot be observed within this world totally composed from its perturbations, and the motion of those perturbations (and thus of any physical body) occurs without any additional 'friction', since any motion is produced by reduction-extension processes of the protofield 'matter' of ether itself. The multiple particular consequences of this picture for the consistent 'unification of interactions' at the 'micro-scale' and causally extended cosmology at the 'macro-scale'

---

system of majority voting is the best and very efficient way to stop any progress in any field (and not only in science). Therefore, irrespective of the well-organised system of imposed 'opinions' of the unitary system of power in science and beyond, today the civilisation *objectively* finds itself before the crucial, and *irreducible*, choice among the dominating system of 'voted truth', leading to the last phase of decadence and the following inevitable demolition, and a new, 'distributed' (non-unitary) type of dynamical organisation of science and society permitting to every person and idea to realise the maximum of its intrinsic possibilities. However, the transition to such qualitatively superior level of organisation, being quite realistic, should necessary involve itself the extended, explicitly multivalued (non-unitary) thinking of the universal science of complexity (see ref. [7] for more details).

outlined in this work (see also refs. [7,11]) are deprived now from the irreducibly large (and always growing) ambiguity of the speculative unitary versions within the canonical science.

Turning now to the *practical* aspect of urgency of transition to the *unreduced*, qualitatively extended approach of the universal science of complexity in fundamental physics and the system of knowledge in general, we note that one of the definite signs of the current crisis of the 'end of science' is that further advancement of knowledge is blocked not only at its fundamental levels, but also in many practical directions involving irreducibly complex phenomena and naturally coming at the foreground of the general progress of civilisation. Indeed, the modern apparently prosperous technical development is persistently limited to a smaller part of the world and to certain fields, directions and way of growth that can support a mechanistically based, artificially simplified approach of the unitary, effectively one-dimensional thinking. Such are all those 'fantastic' computers, energy sources, 'production lines', etc. But the intrinsic, really interesting possibilities of such kind of essentially non-complex applications are severely limited from above and are already basically exhausted, despite a sufficiently large inertia of their external achievements. The irreducible, and very high, dynamic complexity of the real world (including the developing human consciousness at its highest levels) inevitably demands to pass to the corresponding qualitatively different approach of the universal science of complexity, starting already from the simplest, 'physical' problems, such as new, 'sustainable' energy sources, 'non-destructive' interaction with the natural environment ('ecological problems') and real understanding/mastery of its evolution, new types of materials with evidently complex-dynamical behaviour (high-temperature superconductors, all effects of 'strong electron interaction', biochemical systems and 'soft' solids, etc.). Those problems has already become the key subjects of research in many most important practical domains, but their irreducible dynamic complexity is recognised only at an empirical, intuitive level, and the actual approaches used are still totally within the dominating single-valued, unitary paradigm and the related mechanistic 'way of thinking', which excludes any genuine, intrinsically complete understanding and thus efficient applications. As a result, one observes today in all those 'hot' fields a huge accumulation of 'promising' empirical data, but the total absence of a true progress despite really big investments, which is not just 'temporary difficulties': there is no chance to understand and successfully use a hierarchically complex, extremely multivalued phenomenon within an approach supposing and capable to treat only one over-simplified realisation-possibility. The stagnating situation is fields like 'controlled nuclear fusion' and 'high-temperature superconductivity' provides only the most evident examples of this fundamental limit of the unitary approach. The only possible issue from that more hidden, but quite real *crisis in practical domains* consists in transition to the qualitatively new level of the truly adequate, dynamically multivalued description proposed by the *unreduced* concept of dynamic complexity (all the numerous imitations of the genuine complexity within the unitary 'science of complexity', inevitably empirical and inconsistent, can only lead to a deep confusion and further degradation of the situation, cf. [7,12]). The intrinsically complete knowledge of the unreduced science of complexity naturally starts from the most fundamental levels of the universal hierarchy of world complexity described in this work (see also [7-11]), and their adequate understanding can already give, as we have seen above (section 3), the important practical consequences. It is important to emphasize, however, that contrary to the canonical science, composed from basically separated domains of research, the universal science of complexity produces 'simultaneously' the whole hierarchy of complexity of the world in the form of the natural 'complexity development process' [7], and therefore we quickly pass from the lowest to higher levels of complexity and the related applications by using essentially the same concepts and obtaining similar, universally valid results (this is the 'holographic' property of the intrinsically holistic knowledge of the science of complexity [7]). In this way one can see that the same approach that naturally resolves bundles of fundamental problems in micro-physics, will naturally lead to the proper, intrinsically complete solution of the accumulated practical problems from higher complexity levels of ecology, geophysics, biology and medicine, economy, sociology, and the civilisation development in general [7]. Contrary to this, the existing 'mega-projects' of the unitary science, conceived formally for the same purpose, are *basically* incapable to produce the solution, which is confirmed by their actual status, and are maintained instead by financial interests of their participants transforming those pompously advertised 'initiatives' into a huge system of 'scientifically looking' frauds that artificially and egoistically divert the whole amount of the available efforts from the search for the intrinsically complete, and thus the only practically efficient, complex-dynamical solutions. In the meanwhile, the practical difficulties created by the irreducibly complex processes of interaction within the technically developed civilisation quickly grow in quantity, but especially in quality, acquiring a global and irreversible character. The indispensable, and quite evident, 'criterion of truth' remains, however, always the same: it is the *intrinsically complete*, detailed and transparent *understanding* that cannot be replaced with any technical manipulations of the 'advanced empiricism', and the universal science of complexity simply provides the ultimate, rigorously expressed confirmation of this unique genuine motivation for scientific activity, *actually* maintaining its course since the times of ancient civilisations, despite all the *externally* dominating mechanistic inclinations.

The modern urgency to pass to explicit domination of the extended, intrinsically complete way of knowledge development within the unreduced concept of complexity has also a related practical aspect involving very serious, global *dangers of scientific research* not occasionally becoming quite real and potentially fatal just at the end of the unitary science. Those problems are rather clearly 'felt' today at the intuitive and empirical level, but the existing estimates and practical conclusions have inevitably a poorly substantiated, and therefore rather uncertain character. Here again the unreduced science of complexity provides a clear, conscious explanation of the situation, actually confirming the vague 'feelings' about the growing potential danger of (unitary) science. Namely, at the current, fundamentally inevitable 'end' of its development, the single-valued science has exhausted its actually very limited possibilities for consistent description of the intrinsically multivalued reality, but has attained, at the same time, the maximum, and actually quite elevated, level of purely quantitative, mechanical power of its various tools which typically starts exceeding, in the modern epoch, the respective characteristic values of world complexity in practically all kind of system. The 'method' of the unitary science remaining as fundamentally empirical as ever is therefore reduced today to *efficient* destruction of multiple complex-dynamical 'links' in the universal hierarchy of world complexity as a result of erratic 'shooting in all directions' (the irreplaceable 'trial-and error' technique of the unitary empiricism), which can lead to unpredictable consequences of further arbitrary recombination or even fatal disappearance of those links. The particularity of modern epoch is in the fact that before the same chaotic 'shooting' of the canonical empiricism had typically insufficient power for basic links destruction, and therefore the 'alchemy' of the 'trial-and-error' method of the mechanistic approach had poor consequences rather for the efficiency of research itself (even though the 'feeling' of danger from science was often even stronger than today, which demonstrates the insufficiency of conclusions based on intuitive fears alone).

An important particular example of such potentially dangerous application of the 'stupid' high power of the 'developed' unitary machine, directly related to the results of this work, concerns experimental research in high-energy physics. As we have seen in section 3, the really first-principle, causally complete picture of the world shows that the highest value of mass-energy-complexity of an elementary object characterising also the 'strength' of the 'double-layer' ('sandwich') construction of the world is around the highest already observed mass, 100 GeV, which is also the order of magnitude of the renormalised 'Planck unit of mass' determining the observable limits of that construction. At the same time, the purely mechanistic, 'dimensional' estimates of the unitary approach give $10^{19}$ GeV for the Planck mass, leaving additional 17 orders of magnitude of 'free space' for hypothetical 'new species' of particles, which is readily filled up with all kind of (basically unlimited) 'fantasies' of the unitary 'mathematical physics'. The experimental strategy ensuing from this canonical picture is reduced precisely to the described powerful 'shooting' in search of *mathematically* 'predicted' (but actually non-existing) particles with energies that *should*, and today increasingly can, considerably exceed the *real* 'strength of the wall', 100 GeV. This situation can be figuratively, but correctly, compared to a community living within a well isolated building and trying to shoot at the wall from inside using very powerful and expensive weapons, 'just by curiosity'. We should also take into account that we do not know not only what exists 'outside' our wall of interacting protofields (and especially 'behind' its hidden, gravitational 'sheet' at which properly we shoot with the accelerated species) and how it will behave at the 'direct' contact with our world, but also what can be the behaviour of the disrupted 'wall' itself, taking into account that the 'wall' and the 'outside' are necessarily also complex-dynamical, 'living', and therefore unpredictable entities. It can well happen that the wall is strong or thick enough or at least properly and quickly 'closes' after any 'hole' in it is produced, which seems to be confirmed by existence of very energetic particles within the 'cosmic rays'. However, one can never be sure, since the detailed conditions of our artificial experiments having no direct natural prototype can be quite different from isolated collisions of cosmic radiation, and various known, but unexplained processes of extra-high energetic emission in the universe could provide an example of the effect of a less successful closure of a 'hole' in the 'wall of the world'...

Another characteristic example of the same kind of danger of the blind unitary empiricism can be found at a much higher level of complexity, in the modern genetic research and 'human genome project' in particular. The complexity here is measured not in the units of energy but rather directly in realisation numbers, and we have the same characteristic situation when the unitary approach can *technically* manipulate at the characteristic level of complexity, by arbitrary transposing the genetic material, but without the slightest idea about the *complete*, unreduced operation of the *complex-dynamical* system in question. Already the elementary considerations within the universal science of complexity show that the modern 'evolutionary genetics' and other related fields of the canonical science are very far from the real dynamic complexity of the phenomena occurring within the *unreduced* dynamics of *interacting* genes [7] (the approach of the unitary science to this interaction is generally very similar to the typical *logical trap* of perturbation theory in physics, where the initial assumption about a small effect of interaction is 'self-consistently' confirmed by the obtained results, which is simply equivalent to *throwing out* any other cases). The danger of the mechanistic manipulations with genes is evident: (actually) arbitrary manipulations over unexplained complex system can give uncontrollable behaviour of the results, sometimes — and this is especially dangerous — accompanied with the *illusion* of the mechanistic 'total control' over the 'whole system' reduced in reality to the vanishingly small part of reality accessible to the unitary thinking.

Whatever are the details and reality of various dangers of this kind, there is always the same, unique 'right solution' coinciding with the above 'criterion of truth' and the related universal 'ethical principle' in science: every research is possible and desirable, but *only if* it is accompanied by the *intrinsically complete understanding* that *can* indeed be provided, as we have *explicitly* shown, by the unreduced concept of complexity based on the naturally emerging dynamic redundance of physically real entities. In this way the universal science of complexity avoids the 'irresolvable' contradiction of the canonical science between the desirable *completeness* of knowledge and equally necessary potentiality of its *permanent development*: the internally continuous knowledge of the science of complexity is as permanently developing as the world it *adequately* describes (being its inherent constituent), but is also always ultimately complete at *each* attained level (as a complex-dynamical 'gas' always taking the 'whole accessible volume'). It is this extended character and permanently *growing* complexity of the new knowledge that provide its decisive advantage over the invariably dull, repetitive mechanistic manipulations of the single-valued, zero-complexity thinking, totally exhausted now. The necessary transition to the explicitly complex, dynamically multivalued thinking not only provides the unique opportunity to avoid the serious dangers of blind mechanistic manipulations with the living reality, but gives one the highest possible, and basically inexhaustible, *pleasure* of unceasing, conscious creation within the naturally unified, but always growing diversity of being. The results described in this work and concerning the lowest observed levels of this universal hierarchy of complexity will hopefully provide a convincing confirmation of both conceptual consistency and practical utility of the proposed new approach.

## Instead of acknowledgement

The creation of this work was absolutely impossible within the support the author has from the National Academy of Sciences of the free, democratic Ukraine which uses its freedom for democratically paying to the author, a 'senior scientist', the unpredictable mean between 0 and 35 US$ per month, after 20 years of work in the same establishment. Nevertheless, the research has been accomplished, eventually due to an exceptional *chance*.

The publication of this work, or any its part, is impossible in any printed 'scientific' edition, since all of them are controlled by well-organised clans of 'high priests' of the unitary science that cannot provide any consistent understanding at all of the physical essence of any of the entities it pretends to describe and which were endowed with such understanding within the present approach. Nevertheless, the book containing extensive presentation of the approach and its results is published [7], due to a pure *chance*, of course.

The presentation of this work in the *exceptionally* liberal Los-Alamos archives could hardly be possible from a place where even the regular electric power supply is not guaranteed, to say nothing about computers, special software, and access to internet. Nonetheless, the world of pure *chance* is full of *miracles*, and the work is presented in the archives and accessible for the largest possible audience, together with other related articles.

Therefore the author would like to express his sincere and unlimited gratitude to Her Majesty the Pure Chance, but since the author has created himself the theory of *causal* randomness presented in this work and proving that even the pure chance always has a well defined origin, he would like especially to express his gratitude to those Creators of Destiny that turned the incredible series of chances in his favour, so that he may forget about the alternative series of chances known as 'ordinary life'.

The author feels also he should express his gratitude to chief bodies of numerous organisations related to science all over the world, including many 'leading' scientists, various memorial, charitable, 'philanthropic', humanitarian, and 'humanistic' foundations, associations and societies, uncountable national and 'international', especially European, 'commissions', 'cooperation initiatives', 'unification funds', 'academies', 'advanced study institutes', conference committees, and other 'elitary circles' — to all of them for *not* providing this research with any support at all and *not* paying any attention to its results (although many of them had the information about it), which has permitted the author to feel himself free from any obligations before any official or unofficial dogmata and deal only with the truth deprived from any 'influences' and 'relations'. (N.B. The author cannot be sure that the parasitic communities consuming each year many milliards of dollars taken from unaware society and spent for 'research' that does not, and actually cannot, propose any consistent solution in principle will be provided with a real gratitude from the forces of providence, but this kind of choice should always remain strictly individual.)